\documentclass[%
 reprint,
 amsmath,amssymb,
 aps,
 longbibliography, 
]{revtex4-2}
\usepackage{array}     
\usepackage{subcaption}
\usepackage{booktabs}
\usepackage{caption}
\usepackage{graphicx}
\usepackage{dcolumn}
\usepackage{bm}
\usepackage{caption} 

\usepackage{subcaption} 
\begin{document}

\preprint{APS/123-QED}

\title{Tabular Prior-data Fitted Network (TabPFN) optimization of residual correction of $\alpha$-decay energy based on Weizsäcker-Skyrme 4 model}

\author{Guimei Long}

\author{Enxi Dong}

\author{Panpan Qi}

\author{Qingning Yuan}


\author{Anqi Yang}
\author{Gongming Yu}
  \email{ygmanan@kmu.edu.cn}
\affiliation{%
College of Physics and Technology, Kunming University, Kunming 650214, China
}%

\author{Haitao Yang}
  \email{yanghaitao205@163.com}
\affiliation{%
School of Physics and Information Engineering, Zhaotong University, Zhaotong 657000, China
}%

\author{Zhangyan Li}
 \email{20230001@ztu.edu.cn}
\affiliation{%
School of Physics and Information Engineering, Zhaotong University, Zhaotong 657000, China
}%

\author{Yanbing Cai}
  \email{yanbingcai@mail.gufe.edu.cn}
\affiliation{%
Key Laboratory of Economic System Simulation of Guizhou Province, Guizhou University of Finance and Economics, Guiyang 550025, China
}%



\begin{abstract}

A physics-guided hybrid framework combining the Weizsäcker–Skyrme-4 (WS4) macroscopic–microscopic mass model with the Tabular Prior-data Fitted Network (TabPFN) is developed to improve the accuracy of $\alpha$-decay energy ($Q_\alpha$) predictions for superheavy nuclei. TabPFN learns the residuals between experimental data and WS4 predictions using nuclear-structure features. Through ablation studies of 18 feature combinations, the optimal feature set is identified, reducing the overall root-mean-square error (RMSE) from 0.2743 to 0.1442 MeV, an improvement of 47.43$\%$. The corrected model improves prediction accuracy across all four odd–even nuclear configurations while preserving shell effects and odd–even staggering patterns along the Po, Rn, and U isotopic chains. The framework is further applied to predict $Q_\alpha$ values for experimentally unmeasured nuclei along the $Z=118–120$ isotopic chains. The extrapolated results retain the overall trends of WS4 and exhibit signatures associated with the predicted neutron shell closure at $N=184$. These findings highlight the potential of physics-guided residual learning for predicting the decay properties of superheavy nuclei.

\end{abstract}

\maketitle


\section{Introduction}

$\alpha$ decay is one of the principal decay modes of radioactive nuclei and has been extensively investigated since the pioneering studies of Rutherford \cite{1,2}. In the early 20th century, Geiger and Nuttall established the well-known Geiger–Nuttall law, which reveals an empirical relationship between the $\alpha$-decay half-life and decay energy, providing a foundation for quantitative studies of $\alpha$ decay \cite{3,4}. For superheavy nuclei, $\alpha$ decay is a dominant decay channel and plays a crucial role in identifying new nuclides, probing nuclear structure, and confirming the synthesis of new elements \cite{5,6,7,8}. The $\alpha$-decay energy, $Q_\alpha$, is a key physical quantity in the study of heavy and superheavy nuclei. According to the Geiger–Nuttall law, the $\alpha$-decay half-life exhibits an exponential dependence on $Q_\alpha$, such that even small uncertainties in the predicted decay energy can lead to substantial deviations in the estimated half-life \cite{9,10,11,12}. Therefore, accurate predictions of $Q_\alpha$ are essential for understanding nuclear structure, evaluating the reliability of nuclear mass models, and exploring the decay properties of experimentally unknown nuclei \cite{13,14,15,16,17,18}.

Various theoretical nuclear mass models have been developed to predict the $\alpha$-decay energies of superheavy nuclei. Macroscopic–microscopic models, such as the finite-range droplet model (FRDM) \cite{15,19,20} and the Weizsäcker–Skyrme-4 (WS4) model \cite{21}, combine macroscopic liquid-drop descriptions with microscopic corrections that account for shell and pairing effects. Self-consistent mean-field approaches, including relativistic mean-field (RMF) \cite{22}, Skyrme–Hartree–Fock–Bogoliubov (SHFB) \cite{23}, and Hartree–Fock–Bogoliubov (HFB) models \cite{24}, describe nuclear properties through effective nucleon interactions and self-consistent calculations. Semi-empirical mass models, such as the Bhagwat model \cite{25,26,27}, provide an alternative approach to nuclear mass predictions. Among these models, WS4 exhibits high predictive accuracy for nuclear masses and $\alpha$-decay energies, particularly in the heavy and superheavy regions, while incorporating important nuclear shell effects. These features make WS4 a suitable theoretical baseline for the machine-learning-based residual correction developed in this work \cite{28}.

The macroscopic-microscopic mass model Weizs\"acker –Skyrme-4 (WS4) combines liquid-drop contributions with Strutinsky shell corrections, pairing correlations, and Wigner corrections to describe both the macroscopic properties and microscopic structure of the atomic nuclei. It provides accurate predictions of ground-state nuclear masses across a broad range of nuclides, including superheavy nuclei, at relatively low computational cost, making it a widely used benchmark for systematic studies of $\alpha$-decay energies \cite{29,30,31}. Despite its overall accuracy, WS4 exhibits systematic deviations in regions near shell closures, rapid deformation transitions, and neutron-rich superheavy nuclei, where complex shell–deformation coupling effects are difficult to capture through globally fitted model parameters \cite{32}. These deviations contain systematic components associated with local nuclear-structure effects rather than purely random fluctuations, suggesting that machine-learning-based residual correction could further improve the predictive accuracy of WS4 \cite{33,34}.

In recent years, machine learning (ML) has been increasingly applied in nuclear physics, including nuclear mass predictions, nuclear radius calculations, and decay-property studies \cite{35,36,37}. Its ability to capture complex nonlinear correlations in nuclear datasets offers considerable potential for improving the predictive accuracy of nuclear properties \cite{38,39}. However, many existing ML approaches to $\alpha$-decay energy prediction rely primarily on direct data-driven fitting without explicitly incorporating the underlying physical mechanisms \cite{40}. Although these approaches can achieve improved accuracy, their physical interpretability remains limited, particularly in explaining the contributions of shell effects and deformation coupling to systematic deviations in theoretical predictions \cite{41}. Therefore, developing ML frameworks that balance predictive accuracy with physical interpretability remains an important challenge in $\alpha$-decay studies \cite{42,43}. Commonly employed ML algorithms include artificial neural networks (ANNs) \cite{44}, support vector machines (SVMs) \cite{45}, random forests (RFs) \cite{46}, and extreme gradient boosting (XGBoost) \cite{47}. These methods generally require training on task-specific datasets and often involve hyperparameter optimization. However, experimental data for heavy and superheavy nuclei remain limited, typically comprising only hundreds of samples. Under such small-sample conditions, conventional ML approaches may be susceptible to overfitting, while hyperparameter optimization can introduce additional uncertainty in model selection and predictive performance. These challenges motivate the exploration of alternative ML methods better suited to limited nuclear datasets.

The Tabular Prior-data Fitted Network (TabPFN), a Transformer-based foundation model for tabular data, offers a promising approach to addressing these challenges. Pretrained on large-scale synthetic datasets, TabPFN performs regression through in-context learning without requiring task-specific retraining or fine-tuning. This capability makes it particularly suitable for small-sample applications while reducing the need for extensive hyperparameter optimization \cite{48}. In nuclear physics, TabPFN has been applied to predict $\alpha$-particle preformation factors, demonstrating its potential to capture nuclear-structure correlations associated with shell effects and odd–even staggering \cite{49}. Motivated by these advantages, we employ TabPFN to learn the systematic residuals of WS4 and develop a physics-guided residual correction framework, $\mathrm{WS4}+\mathrm{TabPFN}$, to improve the predictive accuracy of $\alpha$-decay energies.

In this work, we employ the WS4 nuclear mass table developed by Ning Wang, Min Liu, and collaborators to calculate theoretical $\alpha$-decay energies and use TabPFN to learn the residuals between WS4 predictions and experimental values \cite{50}. Rather than directly fitting $\alpha$-decay energies, the proposed framework incorporates physically motivated nuclear-structure features to capture systematic deviations associated with microscopic effects that are not fully described by WS4. To evaluate the contributions of different physical features, we conduct ablation studies involving 18 feature combinations. The results identify \texttt{g17} as the optimal feature set, reducing the overall root-mean-square error (RMSE) by 0.4743 compared with the original WS4 model. Further analyses demonstrate that the residual correction effectively reduces systematic deviations while retaining important nuclear-structure characteristics, including shell effects and odd--even staggering. These findings highlight the potential of the WS4+TabPFN framework for improving $\alpha$-decay energy predictions in heavy and superheavy nuclei\cite{51,52}.

The remainder of this paper is organized as follows. Section II presents the WS4 theoretical model, the residual-learning framework, and the physics-guided feature selection strategy, together with the corresponding ablation studies. Section III evaluates the predictive performance of the $\mathrm{WS4}+\mathrm{TabPFN}$ model through residual analysis, odd–even nuclear configurations, and representative isotopic chains, with particular emphasis on the preservation of shell effects and odd–even staggering. Section IV presents extrapolated $\alpha$-decay energy predictions for the $Z=118–120$ isotopic chains.

\section{THEORETICAL FRAMEWORK}

\subsection{WS4 Theoretical Model }

The WS4 (Weizsäcker--Skyrme-4) model \cite{14} is a global macroscopic--microscopic nuclear mass model. Its total binding energy \(E_{\mathrm{b}}\) is consisted of two parts: the macroscopic liquid-drop energy \(E_{\mathrm{LD}}\) and the microscopic shell correction energy \(E_{\mathrm{mic}}\). For the 2353 known nuclide masses (\(N,Z\ge 8\)) in AME2012, this model gives a global root-mean-square deviation of 298 keV, which is the first time that a mean-field-based macroscopic--microscopic framework has crossed the 0.3 MeV accuracy threshold \cite{21}. Put more concretely, the total binding energy in the WS4 model can be written as the sum of the macroscopic liquid-drop energy and the microscopic shell correction energy,
  \begin{equation}
E_{\mathrm{b}}(A,Z)=E_{\mathrm{LD}}(A,Z)+E_{\mathrm{mic}}(A,Z),
\label{eq:bindenergy}
\end{equation}
here, the macroscopic liquid-drop energy \(E_{\mathrm{LD}}\) is the smooth trend of the nuclear binding energy, while the microscopic shell correction energy \(E_{\mathrm{mic}}\) accounts for the local fluctuations induced by quantum shell effects\cite{53}. The macroscopic liquid-drop energy WS4 can be written as
\begin{equation}
\begin{split}
E_{\mathrm{LD}}(A,Z) &= a_v A + a_s A^{2/3} + E_C \\
&\quad + a_{\mathrm{sym}} I^2 A f_s + a_{\mathrm{pair}} A^{-1/3} \delta_{np} + \Delta w,
\end{split}
\label{eq:LD_energy}
\end{equation}
here, \(a_v=-15.5181\), \(a_s=17.4090\), \(a_{\mathrm{sym}}=30.1594\), and \(a_{\mathrm{pair}}=-5.8166\) are coefficients of the volume energy, surface energy, Coulomb energy, symmetry energy, pairing energy, and Wigner correction, respectively. The Coulomb energy term is qiven by
\begin{equation}
E_C
= a_c \frac{Z^2}{A^{1/3}}
\left(1-0.76Z^{-2/3}\right).
\label{eq:EC}
\end{equation}
The factor \(1-0.76Z^{-2/3}\) in the parentheses is a finite-size correction to the Coulomb energy. If the nucleus were treated as a uniformly charged sphere with point-like protons, the Coulomb energy would simply scale as \(Z^2/A^{1/3}\). However, protons have a finite charge radius, and there is a diffuse layer near the nuclear surface, so the charge distribution is not ideal. That is why this correction has to be included. The symmetry energy is expressed as
\begin{align}
E_a &= a_{\mathrm{sym}}(A,I)\cdot I^2\cdot A\cdot f_s, \label{eq:Ea}\\
I &= \frac{N-Z}{A} \label{eq:I},
\end{align}
where \(I\) is the isospin asymmetry, which measures how unbalanced the numbers of neutrons and protons are inside the nucleus and thus determines how far the nucleus deviates from the \(\beta\)-stability line.

In the macroscopic liquid-drop energy of the WS4 model, the pairing energy \(E_{\mathrm{pair}}\) and the Wigner energy \(\Delta w\) describe, respectively, how nucleon pairing and isospin-symmetry effects contribute to the binding energy. To be sure, both terms are numerically much smaller than the volume or surface energy. That said, they remain indispensable for accurately describing nuclear masses---especially odd-even mass staggering and nuclides in the \(N\approx Z\) region. The pairing energy \(E_{\mathrm{pair}}\) originates from pairing correlations between nucleons: two nucleons of the same kind (proton--proton or neutron--neutron) tend to couple into a pair with zero angular momentum, thereby lowering the energy of the system. In the WS4 model, the pairing energy can be written as
\begin{equation}
E_{\mathrm{pair}}
= a_{\mathrm{pair}} A^{-1/3}\delta_{np},
\label{eq:Epair}
\end{equation}
here, the odd-even correction factor \(\delta_{np}\)  for the four even-odd combinations of nuclides represents the direction and strength of the nucleon pairing effect's contribution to the binding energy. In simple terms, it reflects the odd-even mass staggering phenomenon in nuclei.
\begin{equation}
\label{eq:delta_np}
\delta_{np} = 
\begin{cases}
|I| - I^2, & \text{odd-odd nuclei} \\[1.5ex]
\dfrac{17}{16}(2 - |I| - I^2), & \text{even-even nuclei} \\[1.5ex]
0, & \text{\shortstack[l]{odd-even nuclei \\ or even-odd nuclei}}
\end{cases}
\end{equation}

The physical origin of the Wigner energy is isospin symmetry. When the neutron and proton numbers in a nucleus are nearly equal (i.e., \(N\approx Z\)), the similarity between neutrons and protons under the strong interaction gives rise to extra symmetry correlations that increase the binding energy. In the WS4 model, the Wigner correction term \(\Delta w\) can be the parameterized form of the WS-series models,
\begin{equation}
\label{eq:delta_w}
\Delta w = c_W |I|,
\end{equation}
where \(c_W=0.8705\,\mathrm{MeV}\).

In the WS4 model, the macroscopic liquid-drop energy \(E_{\mathrm{LD}}\) is the smooth trend of the nuclear binding energy. It cannot account for the anomalous stability of nuclei at magic numbers\cite{54}. The microscopic shell correction energy \(E_{\mathrm{mic}}\), by contrast, provides a quantitative handle on this ``extra'' stability---it comes from quantum oscillation effects that arise as nucleons fill quantum levels.

Shell correction starts from solving the motion of nucleons in the mean field. The WS4 model uses an axially deformed Woods--Saxon potential,

\begin{equation}
V(r)
= -\frac{V_0}{1+\exp\left(\dfrac{r-\mathcal{R}(\theta)}{a}\right)},
\label{eq:WoodsSaxon}
\end{equation}
here the potential depth is modulated by the isospin asymmetry,
\begin{equation}
V_q = V_0 \pm V_s I,
\label{eq:Vq}
\end{equation}
where \(V_s\) is the macroscopic symmetry-energy coefficient \(a_{\mathrm{sym}}\), which ties the microscopic and macroscopic parts together. The plus sign is for neutrons and the minus sign for protons. To properly describe the nucleon spin-orbit splitting, the single-particle potential also includes a spin-orbit term,
\begin{equation}
V_{\mathrm{s.o.}} = -\lambda \left( \frac{\hbar}{2Mc} \right)^2 \nabla V \cdot (\vec{\sigma} \times \vec{p}),
\end{equation}
\begin{equation}
\lambda
= 32\lambda_0\left(1 \pm \frac{1}{3}I - I^2\right),
\label{eq:lambda}
\end{equation}
where the isospin dependence of the spin-orbit strength \(\lambda\) is fixed according to the Skyrme energy density functional, with the plus sign for neutrons and the minus sign for protons; \(M\) is the free nucleon mass, while \(\boldsymbol{\sigma}\) and \(\mathbf{p}\) are the Pauli spin matrix and the nucleon momentum, respectively. The surface diffuseness \(a\) of the Woods-Saxon potential can be written as
\begin{equation}
a = a_0\left(1+2\varepsilon\delta_q\right),
\label{eq:a}
\end{equation}
where, in neutron-rich nuclei (\(I>I_0\)), \(\delta_q=1\) for neutrons and \(\delta_q=0\) for protons; in proton-rich nuclei (\(I<I_0\)), \(\delta_q=1\) for protons and \(\delta_q=0\) for neutrons; otherwise \(\delta_q=0\). The radial coordinate of the nuclear surface is given by
\begin{equation}
\mathcal{R}(\theta)
= c_0 R\left[1+\beta_2 Y_{20}(\theta)
+\beta_4 Y_{40}(\theta)
+\beta_6 Y_{60}(\theta)\right],
\label{eq:Rtheta}
\end{equation}
where \(R=r_0A^{1/3}\) is the single-particle potential radius (taken to be the same for protons and neutrons in a given nucleus), \(c_0\) is a scaling factor determined by the constant-volume condition, which reflects the incompressibility of nuclear matter, and \(Y_{lm}(\theta,\varphi)\) are spherical harmonics. Combining the Woods-Saxon potential above with the potential-depth modulation, the spin-orbit term, and the surface-diffuseness formula, we solve the Schrödinger equation and finally obtain the single-particle levels \(\varepsilon_{q,i}\) of the nucleus. The shell correction is then calculated with the Strutinsky method.

The microscopic shell correction energy in the WS4 model builds on the Strutinsky shell correction and further introduces a surface-diffuseness correction factor and a mirror-nucleus constraint:
\begin{equation}
\Delta E = c_1 f_d E_{\mathrm{sh}} + |I| E'_{\mathrm{sh}},
\label{eq:DeltaE}
\end{equation}
where \(E_{\mathrm{sh}}\) is the shell energy of the target nucleus and \(E'_{\mathrm{sh}}\) is the shell energy of its mirror nucleus. The coefficient \(c_1=0.6309\) is a scaling factor used to correct the systematic overestimation of the shell-correction amplitude by the Strutinsky procedure. \(f_d\) represents the correction of the surface diffuseness to the shell-effect strength. The \(|I|\) term comes from the mirror-nucleus constraint associated with isospin symmetry and can improve the model accuracy by about 10\%. In other hands, WS4 does not treat surface diffuseness merely as an empirical term of the liquid-drop model; instead, it lets this quantity enter both the macroscopic symmetry energy and the microscopic shell structure.

In simple terms, nuclear deformation changes the single-particle level distribution and therefore affects the shell-correction energy. The WS4 model calculates the total energy separately for different deformation configurations and takes the minimum value as the ground-state prediction. The dependence of the macroscopic energy on nuclear deformation is written as
\begin{equation}
\label{eq:ELD}
E_{LD}(\mathrm{def}) = E_{LD}(\mathrm{sph}) \prod_{k} \left( 1 + b_k \beta_k^2 \right),
\end{equation}
where $b_k$ is the coupling coefficient of the $k$-th order deformation to the macroscopic energy and is basically a fitting parameter. It squeezes the deformation dependence of the macroscopic terms---surface, Coulomb, symmetry energies, and so on---into a multiplicative correction relative to the spherical shape. If $b_k>0$, this order of deformation raises the macroscopic energy; if $b_k<0$, it lowers the macroscopic energy. Meanwhile, $\beta_k$ is the $k$-th order deformation parameter.

After considering deformation configurations $\{\beta_k\}$, the total energy is written as
\begin{equation}
E_{\mathrm{tot}}(\beta_k)
= E_{\mathrm{LD}}(\beta_k)+E_{\mathrm{mic}}(\beta_k),
\label{eq:total-energy}
\end{equation}
subsequently, the $Q_\alpha$ value is obtained as
\begin{equation}
\begin{split}
Q_\alpha
&= E_{\mathrm{LD}}(A-4,Z-2)-E_{\mathrm{LD}}(A,Z) \\
&\quad +E_{\mathrm{mic}}(A-4,Z-2)-E_{\mathrm{mic}}(A,Z)
+E_\alpha .
\end{split}
\label{eq:q-alpha}
\end{equation}


\subsection{Residual Learning Framework }

The WS4 model is one of the most accurate macroscopic--microscopic nuclear mass models currently available. In the heavy and superheavy regions, its global root-mean-square (RMS) deviation for nuclear masses is about 298~keV, and its prediction accuracy for $Q_\alpha$ is better than 300~keV \cite{21}. That said, the model still has limitations in describing local shell effects, odd--even effects, and deformation effects---microscopic physics that is not fully resolved. As a result, an observable systematic residual bias remains between theoretical predictions and experimental values. To address this, we introduce a physics-guided residual correction strategy. It keeps the macroscopic physical constraints of WS4 intact while using machine learning to compensate for microscopic structural information that the model does not fully capture.

Let $Q_\alpha^{\mathrm{WS4}}$ is the theoretical decay energy given by the WS4 model, and $Q_\alpha^{\mathrm{exp}}$ the corresponding experimental value obtained from the AME mass excess\cite{55}. Then, the residual can be written as
\begin{equation}
\delta Q_\alpha = Q_\alpha^{\mathrm{exp}} - Q_\alpha^{\mathrm{WS4}}.
\label{eq:residual}
\end{equation}

This residual reflects microscopic physical effects that WS4 does not fully describe, including shell and deformation effects. It is important to note that this residual is not pure random noise. Rather, it contains a systematic component tied to nuclear-structure features, so it can be treated as a learnable correction signal. In practice, we split the data into a training set, a validation set, and an independent prediction set. Machine learning is then used to model the residual: by learning from the training set, we find the functional mapping between the residual and the input features:
\begin{equation}
\delta Q_\alpha = f_{\mathrm{model}}(\mathbf{X}),
\label{eq:mapping}
\end{equation}
where $\mathbf{X}$ is the physics-driven input feature set, and $f_{\mathrm{model}}$ is learned from the training set. The validation set is only passed through the same mapping to check whether the residuals are reduced after correction; it is not involved in any training. The final corrected $Q_\alpha$ is
\begin{equation}
Q_\alpha^{\mathrm{WS4+model}}
= Q_\alpha^{\mathrm{WS4}} + f_{\mathrm{model}}(\mathbf{X}),
\label{eq:corrected}
\end{equation}
where $Q_\alpha^{\mathrm{WS4+model}}$ is the final prediction after residual correction. What this buys us is a residual-correction strategy that combines the strengths of the theoretical model and the data-driven approach. On the one hand, it inherits the macroscopic physical regularities already validated in WS4, which helps avoid the physically unreasonable behavior that pure machine-learning models can easily produce. On the other hand, it uses machine learning to capture shell effects, deformation effects, and other microscopic nuclear-structure information that traditional models struggle to describe analytically. As a result, it improves the prediction accuracy of $\alpha$-decay energies even when heavy-nucleus experimental samples are limited.

Under the residual learning framework, correction performance depends not only on the machine-learning algorithm itself but also on whether the input features can effectively carry microscopic physics that WS4 does not capture. Based on the theoretical output table of the WS4 nuclear mass model, we selected relevant quantities as input features. The mass number \(A\), proton number \(Z\), and neutron number \(N\) serve as basic variables. They describe the overall nucleon configuration and the location on the nuclear chart, and they form the baseline input for capturing global mass trends. \(Z\) and \(N\) directly determine where a nucleus sits on the nuclear chart and can encode trends in macroscopic liquid-drop effects such as volume, surface, Coulomb, and symmetry energies. They can capture global residual variations with nucleon number, but they lack microscopic information such as deformation and shell structure. To address this, we further introduce the shell correction \(E_{\mathrm{sh}}\) and the deformation energy \(E_{\mathrm{def}}\). \(E_{\mathrm{sh}}\) is given by the Strutinsky shell correction in WS4 and directly encodes shell-structure information relative to closed shells. \(E_{\mathrm{def}}\) comes from energy minimization over multidimensional deformation configurations and reflects the contribution of collective deformation degrees of freedom to the binding energy when a nucleus deviates from spherical shape. We also include the quadrupole deformation parameter \(\beta_2\) as a single deformation descriptor, and the odd--even correction factor \(\delta\) to characterize odd--even mass staggering caused by nucleon pairing. In addition, we introduce the distances \(d_Z\) and \(d_N\) from the proton and neutron numbers to the nearest magic numbers, together with the corresponding valence nucleon numbers \(N_p\) and \(N_n\) \cite{56}. These explicitly represent how far a nuclide is from closed shells and its valence-nucleon configuration. Smaller \(d_Z\) and \(d_N\) mean that the nuclide is closer to the corresponding magic closed shell, where shell effects are expected to be more pronounced. \(N_p\) and \(N_n\) further encode how valence nucleons in open-shell regions modulate shell correction and deformation. These features all have clear physical meaning and together form the input feature set for residual learning. This allows the model to retain the macroscopic physical constraints of WS4 while carrying out targeted residual correction for microscopic effects such as deformation, pairing, and shell-transition regions.

To quantitatively identify the contribution of the above information, we designed 18 controlled ablation groups with \((A,N,Z)\) as the common baseline (Table~I). g1 is the baseline; g2--g7 introduce six single factors one at a time; g8--g13 examine pairwise combinations to identify complementarity and redundancy among factors; g14--g18 consist of four three-factor combinations plus the full set, forming leave-one-out checks. This design covers a complete attribution chain from ``single factor'' to ``combination'' to ``leave-one-out.'' It enables the model to retain WS4 macroscopic constraints while specifically compensating for its systematic deficiencies in shell-transition and deformation-transition regions.

\begin{table}[!htbp]
\centering
\caption{Feature set and input descriptors}
(g1 serves as the common baseline; g2–g7 are single-factor groups, g8–g13 are two-factor combination groups, and g14–g18 are three-factor combination and full-feature groups.
)
\label{tab:feature}
\begin{tabular}{ll}
\hline
Feature set & input descriptors \\
\hline
g1  & \([A,\ N,\ Z]\) \\
g2  & \([A,\ N,\ Z,\ \delta]\) \\
g3  & \([A,\ N,\ Z,\ E_{\rm def}]\) \\
g4  & \([A,\ N,\ Z,\ \beta2]\) \\
g5  & \([A,\ N,\ Z,\ E_{\rm sh}]\) \\
g6  & \([A,\ N,\ Z,\ d_z,\ d_N]\) \\
g7  & \([A,\ N,\ Z,\ N_{\rm p},\ N_{\rm n}]\) \\
g8  & \([A,\ N,\ Z,\ E_{\rm def},\ d_z,\ d_N]\) \\
g9  & \([A,\ N,\ Z,\ E_{\rm def},\ E_{\rm sh}]\) \\
g10 & \([A,\ N,\ Z,\ E_{\rm def},\ N_{\rm p},\ N_{\rm n}]\) \\
g11 & \([A,\ N,\ Z,\ E_{\rm sh},\ d_z,\ d_N]\) \\
g12 & \([A,\ N,\ Z,\ E_{\rm sh},\ N_{\rm p},\ N_{\rm n}]\) \\
g13 & \([A,\ N,\ Z,\ d_z,\ d_N,\ N_{\rm p},\ N_{\rm n}]\) \\
g14 & \([A,\ N,\ Z,\ E_{\rm def},\ d_z,\ d_N,\ N_{\rm p},\ N_{\rm n}]\) \\
g15 & \([A,\ N,\ Z,\ E_{\rm def},\ E_{\rm sh},\ d_z,\ d_N]\) \\
g16 & \([A,\ N,\ Z,\ E_{\rm sh},\ d_z,\ d_N,\ N_{\rm p},\ N_{\rm n}]\) \\
g17 & \([A,\ N,\ Z,\ E_{\rm def},\ E_{\rm sh},\ N_{\rm p},\ N_{\rm n}]\) \\
g18 & \([A,\ N,\ Z,\ E_{\rm def},\ E_{\rm sh},\ d_z,\ d_N,\ N_{\rm p},\ N_{\rm n}]\) \\
\hline
\end{tabular}
\end{table}

\subsection{ Machine Learning }

Machine learning provides a data-driven approach to modelling superheavy isotopic chains for which experimental data are currently unavailable, predicting that \(Q_\alpha\) initially increases and then decreases with neutron number \(N\), reaching a peak around \(N = 185\text{–}190\), while nuclides near \(N = 184\) are expected to exhibit enhanced shell stabilization, resulting in a local minimum in \(Q_\alpha\) relative to their neighbouring nuclides \cite{57,58,59,60}. It can dig out hidden nonlinear connections from large datasets and map input features directly to target physical quantities without relying completely on explicit analytical physics formulas. Traditional theoretical models typically build analytical equations based on physical mechanisms. However, some complex microscopic effects are simply too hard to capture fully in theoretical formulas. This leaves us with systematic biases between theoretical calculations and experimental results. So, in nuclear physics, it's common to pair theoretical models with machine learning \cite{28,33,49}. The physics model handles the main prediction, while machine learning steps in to learn the residual---that is, the gap between theory and experiment---and compensates for the microscopic effects the theory misses. This residual learning strategy plays to machine learning's strength in data fitting while keeping the prior constraints of the physical model intact. It is a solid way to balance prediction accuracy with physical interpretability.

In this study, we use TabPFN to modify the WS4 residual. Depending on the proton and neutron parity, we tag the data into four groups: even-even, even-odd, odd-even, and odd-odd. These are stored in separate worksheets. Because too few samples in a single group can easily cause overfitting, we merge all nuclides into a unified dataset for training and validation to make the most of all available samples. Once training is done, we evaluate the correction performance across different parity groups by calculating metrics for each type separately. Our selected nuclide samples cover a proton number \(Z\) range of 63--120 and a neutron number \(N\) range of 84--200, mostly in the heavy and superheavy regions. We split the dataset into training, validation, and extrapolation prediction sets, containing 552, 143, and 78 samples, respectively. The validation set is strictly kept out of the training process. It is only used for ablation comparisons, selecting the best feature combination, and evaluating generalization, which ensures our validation metrics objectively reflect the model's true predictive power. The independent prediction set consists of superheavy nuclides with \(Z = 118\)--\(120\). It has zero overlap with the training and validation sets, and none of these nuclides have experimental decay energy data. This set is reserved for final extrapolation predictions.

Once the features were built and the dataset was split, we ran residual correction experiments on the merged unified dataset for all 18 feature combinations (Table~I) one by one. To keep the comparison between groups fair, we did a fair amount of tuning and ended up sharing a single fixed set of hyperparameters across all ablation experiments to control overfitting (number of base networks \(N_{\mathrm{estimators}} = 8\), Softmax temperature \(\tau = 4.5\)). Meanwhile, to smooth out the statistical fluctuations that come from a single training run, we repeated each experiment under three random seeds (42, 7, and 2024). The feature composition and performance on the training, validation, and overall sets for each group are summarized in Table~II. The inter-group comparison of validation RMSE and the incremental waterfall plot of each group relative to the baseline are shown in Fig.~1(a) and (b), respectively. 

The RMSE is calculated as
\begin{equation}
\mathrm{RMSE} = \sqrt{\frac{1}{N} \sum_{i=1}^{N} \left( Q_{\alpha}^{\mathrm{pred},i} - Q_{\alpha}^{\mathrm{exp},i} \right)^2},
\label{eq:rmse}
\end{equation}
where \(N\) stands for the number of samples, and \(Q_{\alpha}^{\mathrm{pred},i}\) and \(Q_{\alpha}^{\mathrm{exp},i}\) represent the model-predicted and experimental \(\alpha\)-decay energies for the \(i\)-th nuclide, respectively. A smaller RMSE indicates higher prediction accuracy of the model.

\begin{table*}[!t]
\centering
\footnotesize
\renewcommand{\arraystretch}{1.1}
\caption{RMSE of WS4$+$TabPFN across the 18 feature combinations and the relative improvement compared to WS4}
\label{tab:rmse_improvement}
\begin{tabular}{lcccc}
\hline
Model & RMSE (Training set) & RMSE (Testing set) & RMSE (Entire set) & RMSE improvement (\%) \\
\hline
WS4  & 0.2751 & 0.2713 & 0.2743 & ------ \\
g1   & 0.1519 & 0.1830 & 0.1588 & 42.11 \\
g2   & 0.1525 & 0.1838 & 0.1594 & 41.89 \\
g3   & 0.1412 & 0.1763 & 0.1491 & 45.64 \\
g4   & 0.1438 & 0.1842 & 0.1530 & 44.22 \\
g5   & 0.1467 & 0.1756 & 0.1531 & 44.19 \\
g6   & 0.1439 & 0.1762 & 0.1511 & 44.91 \\
g7   & 0.1453 & 0.1747 & 0.1518 & 44.66 \\
g8   & 0.1414 & 0.1748 & 0.1489 & 45.72 \\
g9   & 0.1410 & 0.1736 & 0.1483 & 45.94 \\
g10  & 0.1384 & 0.1711 & 0.1458 & 46.85 \\
g11  & 0.1438 & 0.1745 & 0.1507 & 45.06 \\
g12  & 0.1407 & 0.1700 & 0.1472 & 46.34 \\
g13  & 0.1444 & 0.1750 & 0.1512 & 44.88 \\
g14  & 0.1405 & 0.1727 & 0.1476 & 46.19 \\
g15  & 0.1404 & 0.1712 & 0.1473 & 46.30 \\
g16  & 0.1411 & 0.1710 & 0.1477 & 46.15 \\
g17  & 0.1379 & 0.1689 & 0.1442 & 47.43 \\
g18  & 0.1376 & 0.1691 & 0.1447 & 47.25 \\
\hline
\end{tabular}
\end{table*}


\begin{figure}[htbp]
  \centering
\begin{subfigure}{\linewidth} 
    \centering
    \includegraphics[width=0.9\linewidth]{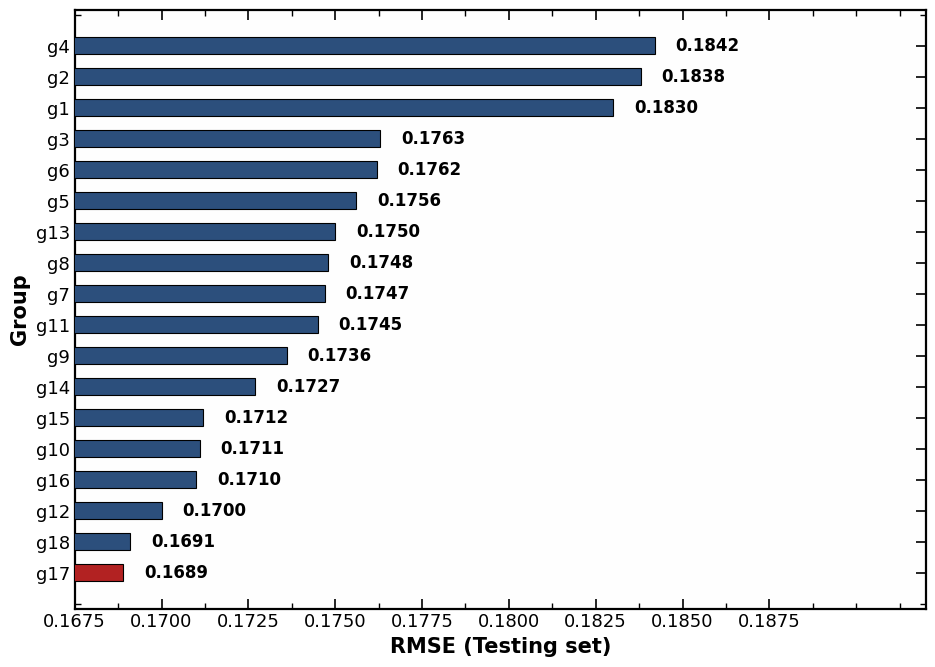}
    \caption{} 
    \label{fig:1a} 
\end{subfigure}

\vspace{0.5cm} 
    \begin{subfigure}{\linewidth}
        \centering
        \includegraphics[width=0.9\linewidth]{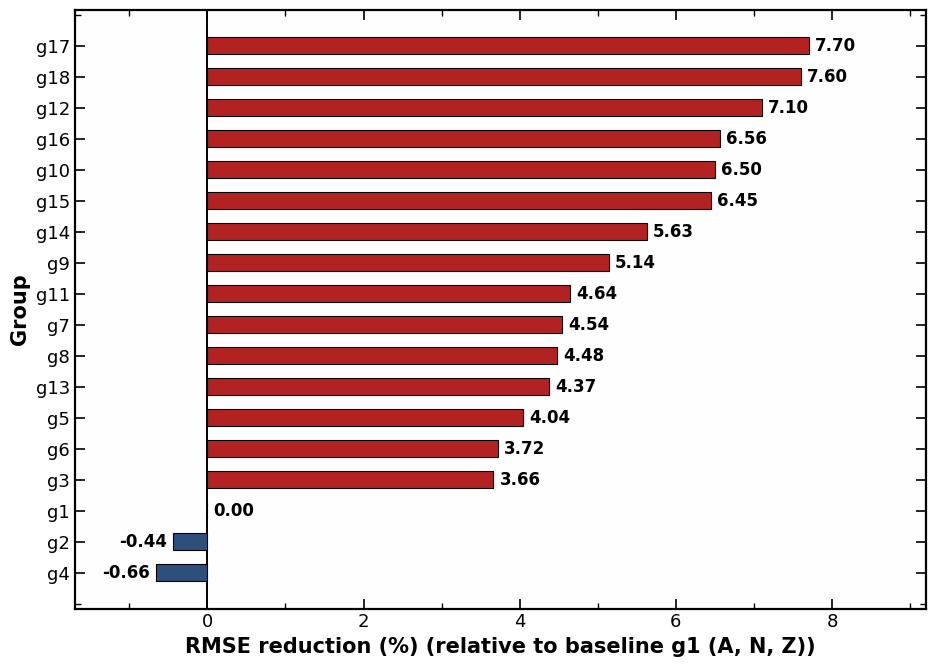}
        \caption{} 
        \label{fig:subfig_b}
    \end{subfigure}
    
\caption{\scriptsize (a) RMSE on the testing set for each feature group; (b) Marginal contribution of each feature group relative to baseline g1 (ablation waterfall).}
  \label{fig:feature-combination-rmse-waterfall}
\end{figure}

As Table~II shows,looking at the overall dataset, WS4 achieves an RMSE of 0.2713 MeV on the validation set. Once calibrated using just the baseline g1 (A, N, Z), it drops down to 0.1588 MeV. This basically tells us that a significant portion of the residual consists of global components that vary smoothly with nucleon numbers, which can be directly captured by the nuclide coordinates. In the single-factor experiments, the deformation energy Edef (g3) yields the biggest gain, suggesting the residual contains deformation information not fully absorbed by WS4. The magic number distances dZ and dN (g6) and valence nucleon numbers Np and Nn (g7) come in second. Interestingly, the pairing label $\delta$ barely contributes anything, and the quadrupole deformation parameter $\beta_2$ (g4) actually leads to a drop in accuracy compared to the baseline features (A, N, Z). This implies that parity information is probably already implicitly encoded within the baseline features.

However, a single deformation parameter isn't enough to capture the nonlinear effects of deformation. In the combination experiments, combining \(E_{\text{def}}\) with \(N_p\) and \(N_n\) (\(g_{10}\)) shows clear complementarity compared to other combos. Moreover, any combination containing \(d_Z\) and \(d_N\) tends to perform worse than the corresponding valence nucleon combo, pointing to information redundancy among magic number distances, shell corrections, and valence nucleons. The leave-one-out test backs this up further: \(g_{17}\), which drops \(d_Z\) and \(d_N\), hits the global optimum and is statistically equivalent to the full-feature \(g_{18}\) within the margin of error, proving that \(d_Z\) and \(d_N\) are replaceable features.

You can also clearly see this in Fig.1(a) and (b) for the validation set: g2 ($\delta$) and g4 ($\beta_2$) show negative increments relative to baseline g1. Edef (g3) has the largest drop among single-factor groups, and g17 shows the deepest cumulative drop relative to g1 across all combinations. Meanwhile, g18 actually drops slightly due to those redundant dZ and dN features. The ranking of all groups perfectly matches the statistical results in Table II . Based on all this, g17 (A, N, Z, Edef, Esh, Np, Nn) is selected as our optimal feature set, and all subsequent experiments are conducted using this set.

\section{\quad RESULTS AND DISCUSSION}

As shown in Fig.~2, we plot the distributions of residuals against the neutron number $N$ for both WS4 (a) and WS4+TabPFN (b). It is pretty obvious that the residuals are scattered within $\pm$0.4 MeV before calibration, but they basically shrink toward the zero line to within $\pm$0.2 MeV after correction. 

Besides, to get a clearer visual of how the residuals behave spatially across the entire heavy-mass region, we generated heatmaps of the mean residuals in the nuclide plane (Fig. 3). Before calibration (a), the residuals display a distinct spatial structure: positive biases dominate in the $N \approx 120\text{--}150$ range, while negative biases pop up in the superheavy region ($Z \gtrsim 100$). After the correction (b), the overall amplitude of the residuals decreases across all regions, and the spatial structure is significantly weakened. Large-scale areas of same-sign bias are basically gone on the nuclide chart. This indicates that TabPFN has effectively absorbed the systematic residuals of WS4.

\begin{figure}[htbp]
  \centering
  \includegraphics[width=1\linewidth]{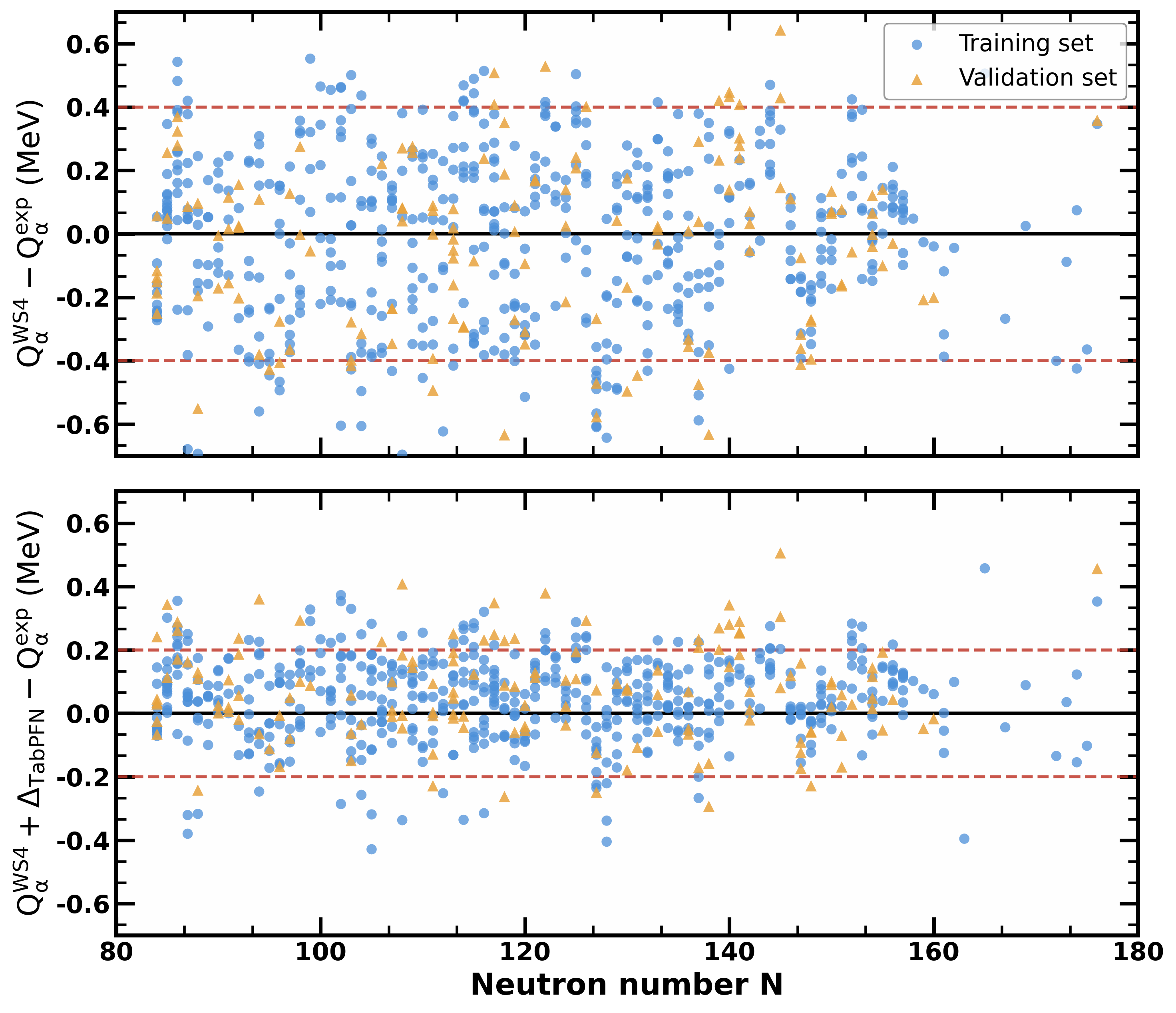}
  \caption{\scriptsize Residual distributions versus neutron number \(N\) for WS4 (a) and WS4+TabPFN (b) (blue dots denote training-set nuclei; red dots denote validation-set nuclei).}
  \label{fig:residual-distribution}
\end{figure}

\begin{figure*}[t]
  \centering
  \includegraphics[width=0.8\linewidth]{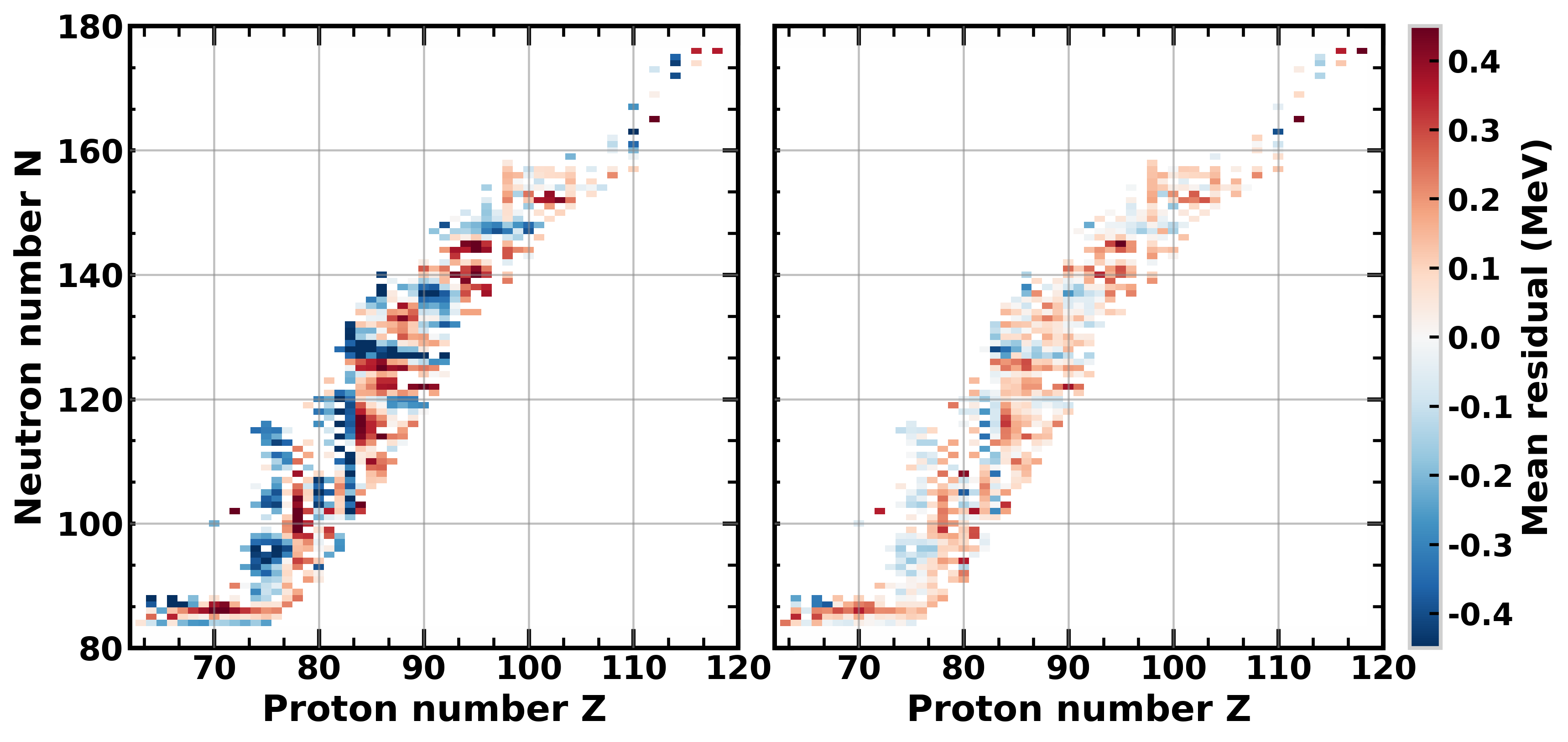}\\[-0.2cm] 
  
  \begin{minipage}{0.4\linewidth}
    \centering (a)
  \end{minipage}%
  \begin{minipage}{0.4\linewidth}
    \centering (b)
  \end{minipage}
  
  \caption{\footnotesize Heatmaps of mean residuals in the nuclide plane: before correction (a) and after correction (b).}
  \label{fig:nuclide-plane-heatmaps}
\end{figure*}

Global metrics show that WS4+TabPFN statistically outperforms the original WS4, but they fail to reveal how the correction behaves across different odd-even configurations. Since these four types of configurations possess distinct pairing mechanisms, their corresponding residual structures are likely to differ as well~\cite{21}. Table III breaks down the RMSE before and after correction across different nuclide types. As we can see, the correction works remarkably well for all four odd-even configurations. Moreover, configurations involving odd nucleons show even greater relative improvements. This suggests that even though the \(\delta\) feature didn't yield any gains in the ablation study, the model still effectively characterizes the pairing-related residual components via the odd-even information implicitly encoded by \((A, N, Z)\).
\begin{table}[htbp]
  \centering
  \caption{Comparison of RMSE before and after correction for the four types of odd-even nuclei.}
  \label{tab:odd-even-rmse}
  \begin{tabular}{ccccccc}
    \toprule
    \textbf{Nucleus} & \textbf{\(N_{\text{train}}\)} & \textbf{\(N_{\text{test}}\)} & \textbf{RMSE} & \textbf{RMSE} & \textbf{RMSE} & \textbf{RMSE} \\
    \textbf{type} & & & \textbf{(WS4} & \textbf{(TabPFN} & \textbf{(WS4} & \textbf{(TabPFN} \\
    & & & \textbf{train)} & \textbf{train)} & \textbf{test)} & \textbf{test)} \\
    \midrule
    even-even & 150 & 38 & 0.2755 & 0.1417 & 0.2860 & 0.1898 \\
    even-odd  & 136 & 37 & 0.2998 & 0.1528 & 0.2815 & 0.1748 \\
    odd-even  & 130 & 35 & 0.2793 & 0.1390 & 0.2434 & 0.1453 \\
    odd-odd   & 136 & 33 & 0.2430 & 0.1149 & 0.2703 & 0.1598 \\
    \bottomrule
  \end{tabular}
\end{table}

\begin{figure*}[!htbp]
    \centering
    \begin{subfigure}{0.32\linewidth}
        \centering
        \includegraphics[width=\linewidth]{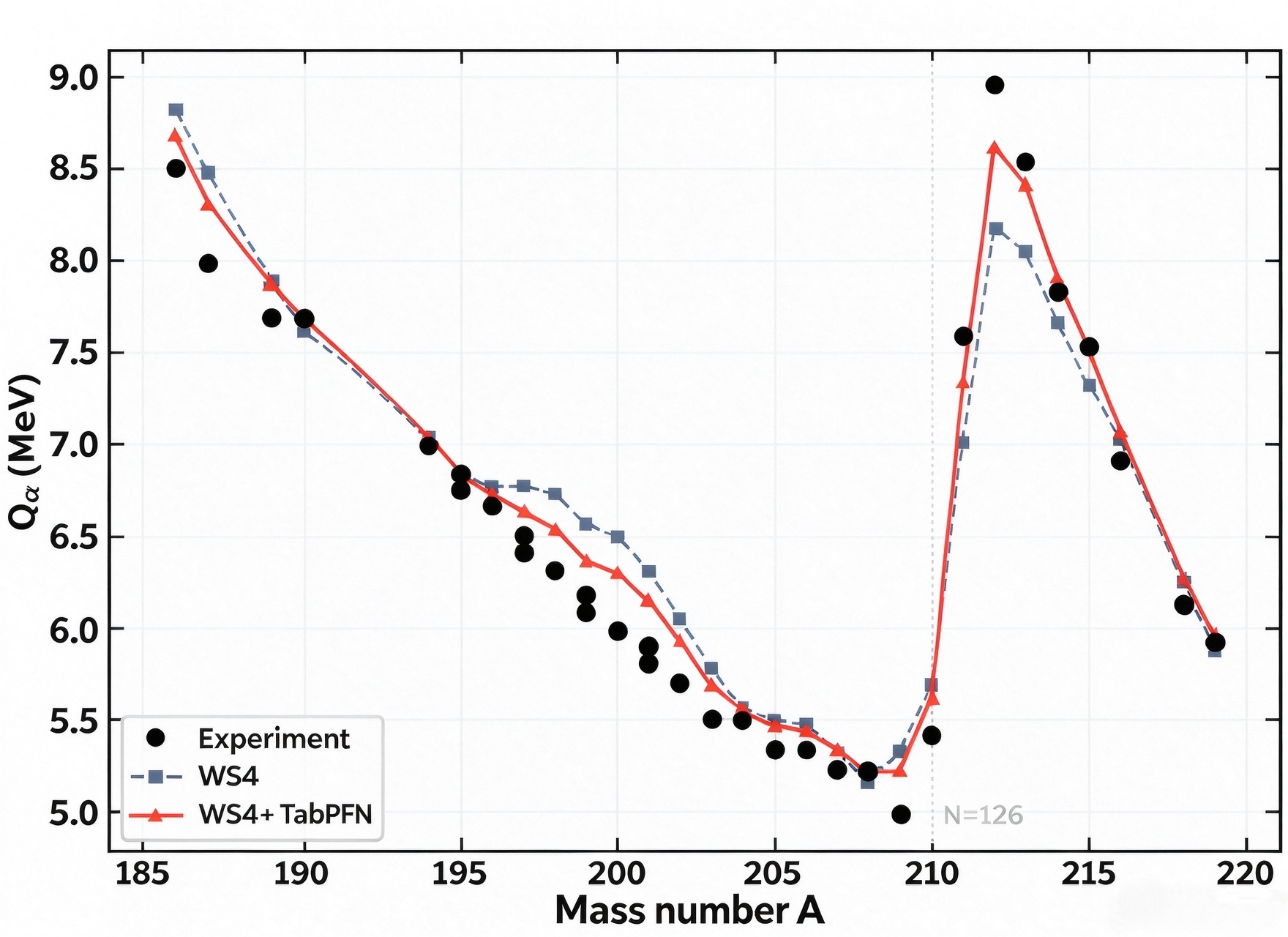}
        \caption{Po isotopic chain (\(Z=84\))}
    \end{subfigure}\hfill
    \begin{subfigure}{0.32\linewidth}
        \centering
        \includegraphics[width=\linewidth]{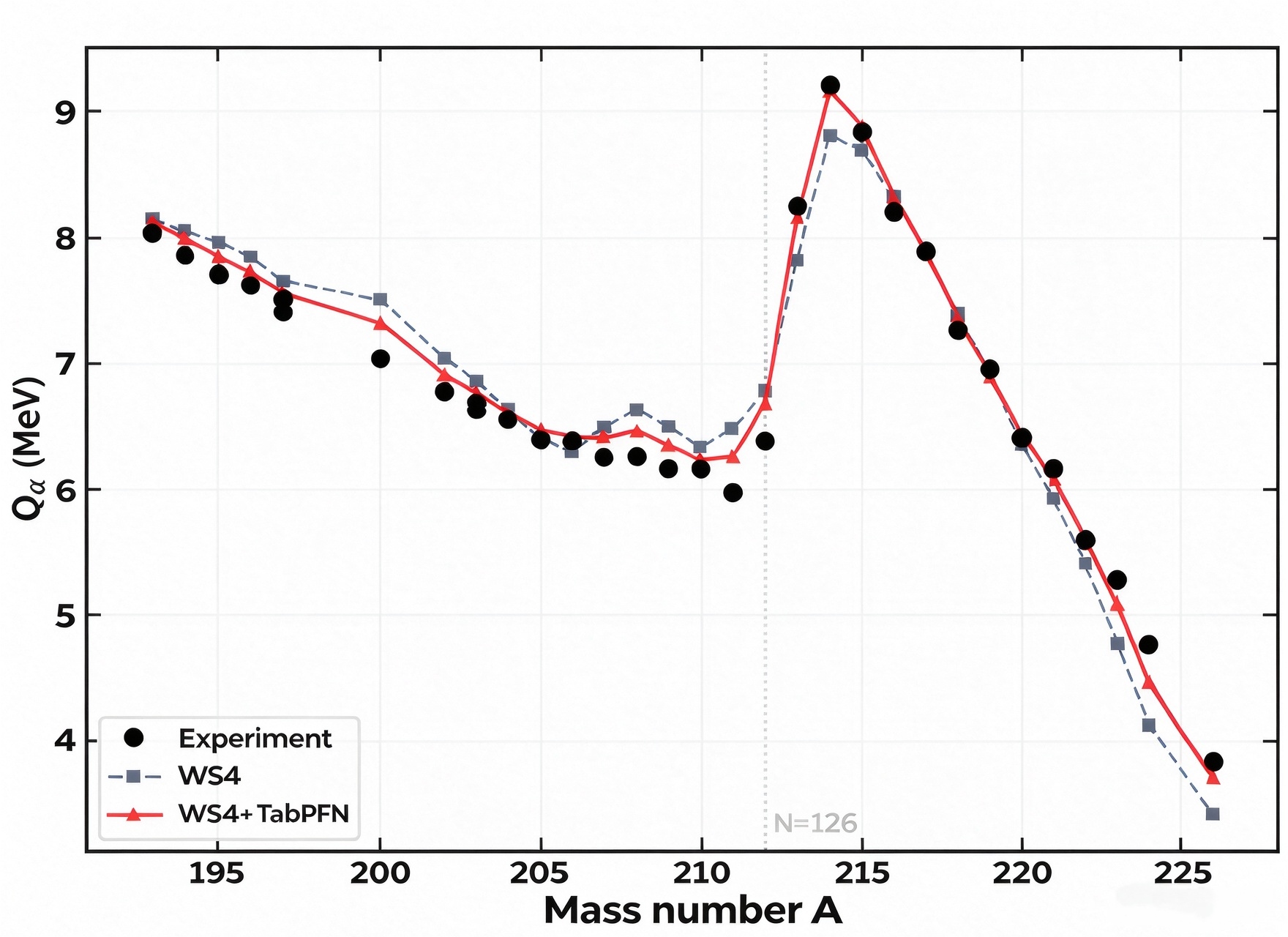}
        \caption{Rn isotopic chain (\(Z=86\))}
    \end{subfigure}\hfill
    \begin{subfigure}{0.32\linewidth}
        \centering
        \includegraphics[width=\linewidth]{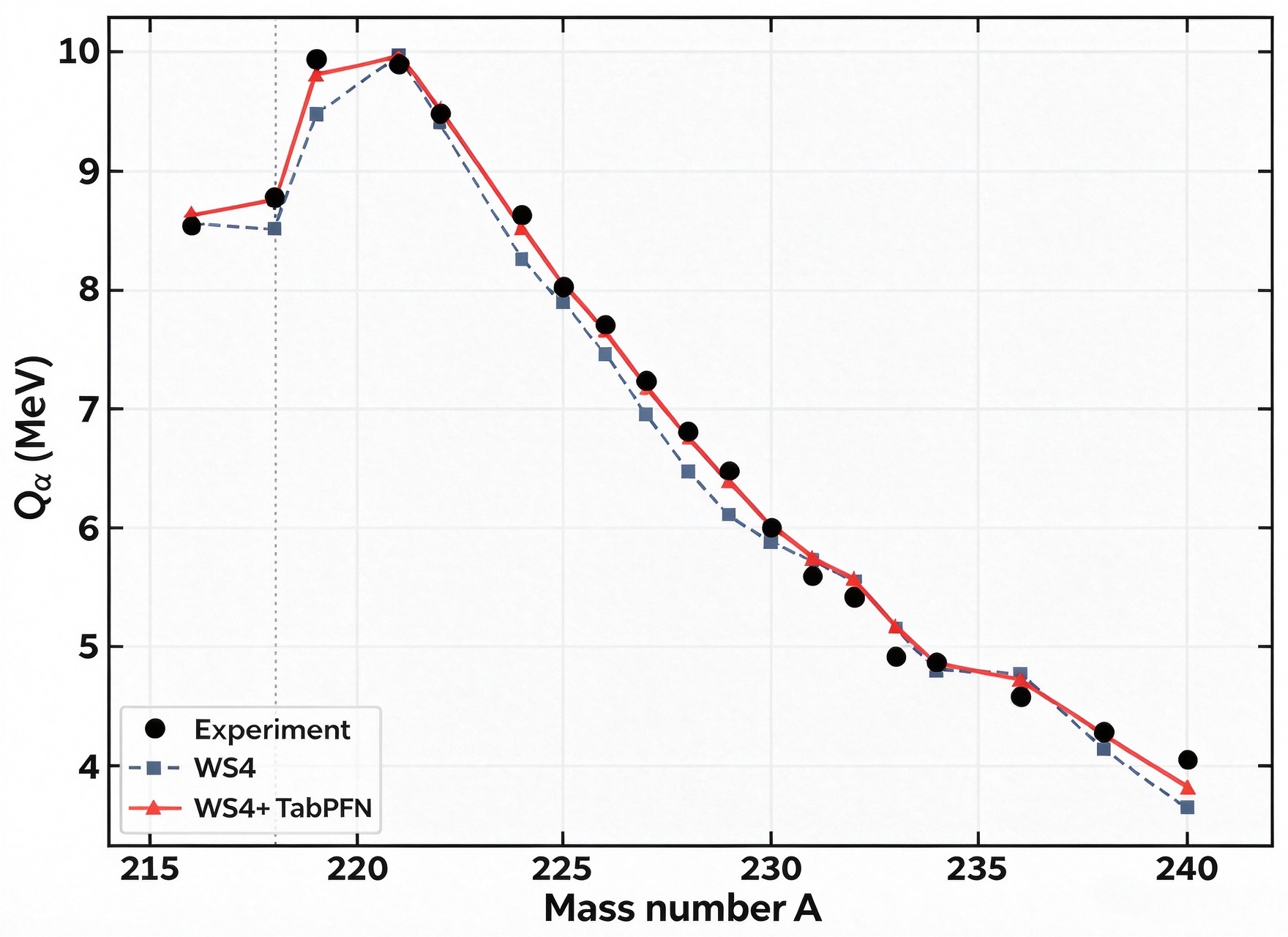}
        \caption{U isotopic chain (\(Z=92\))}
    \end{subfigure}

    \vspace{0.4cm}

    \begin{subfigure}{0.32\linewidth}
        \centering
        \includegraphics[width=\linewidth]{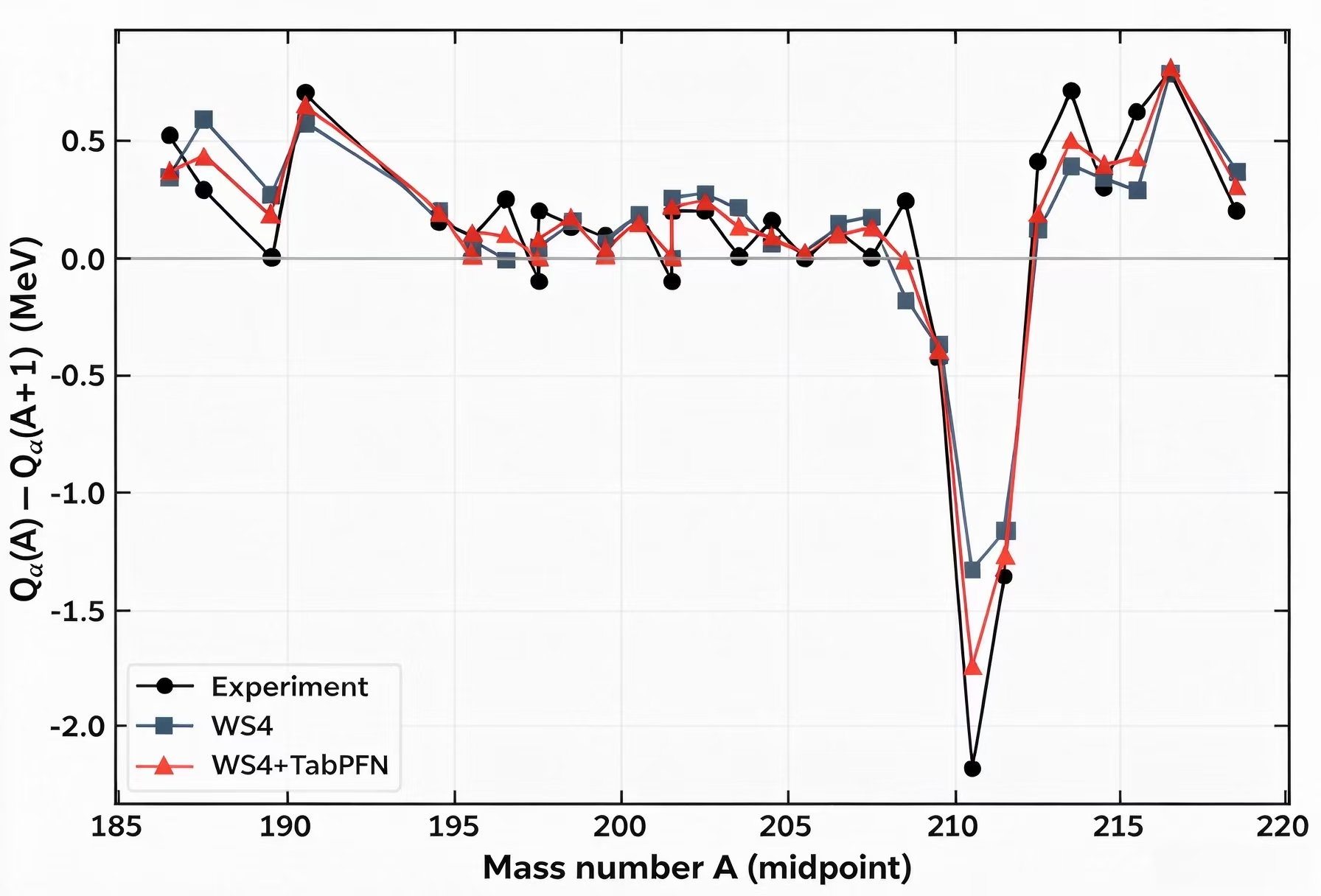}
        \caption{Po isotopic chain (\(Z=84\))} %
    \end{subfigure}\hfill
    \begin{subfigure}{0.32\linewidth}
        \centering
        \includegraphics[width=\linewidth]{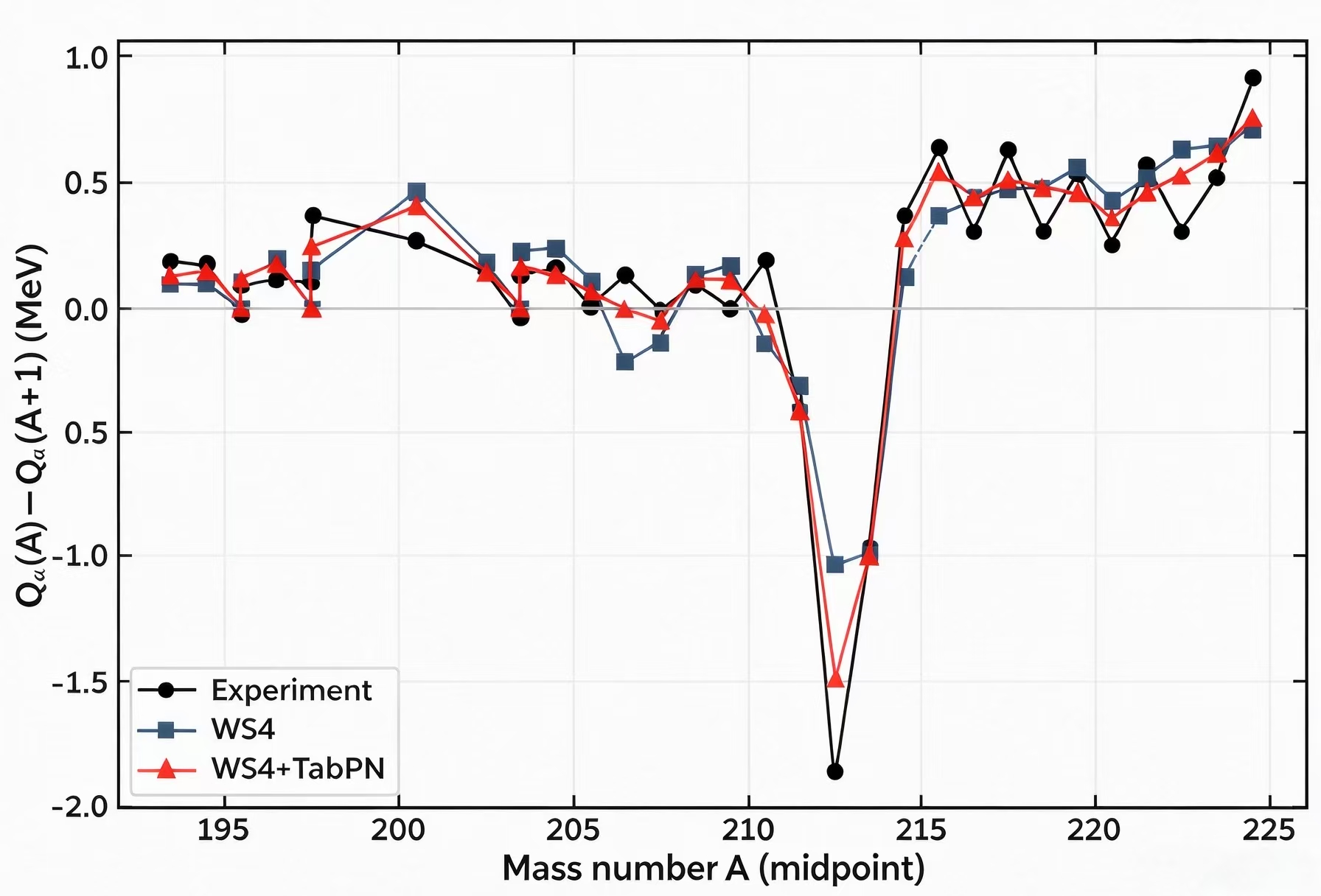}
        \caption{Rn isotopic chain (\(Z=86\)) } 
    \end{subfigure}\hfill
    \begin{subfigure}{0.32\linewidth}
        \centering
        \includegraphics[width=\linewidth]{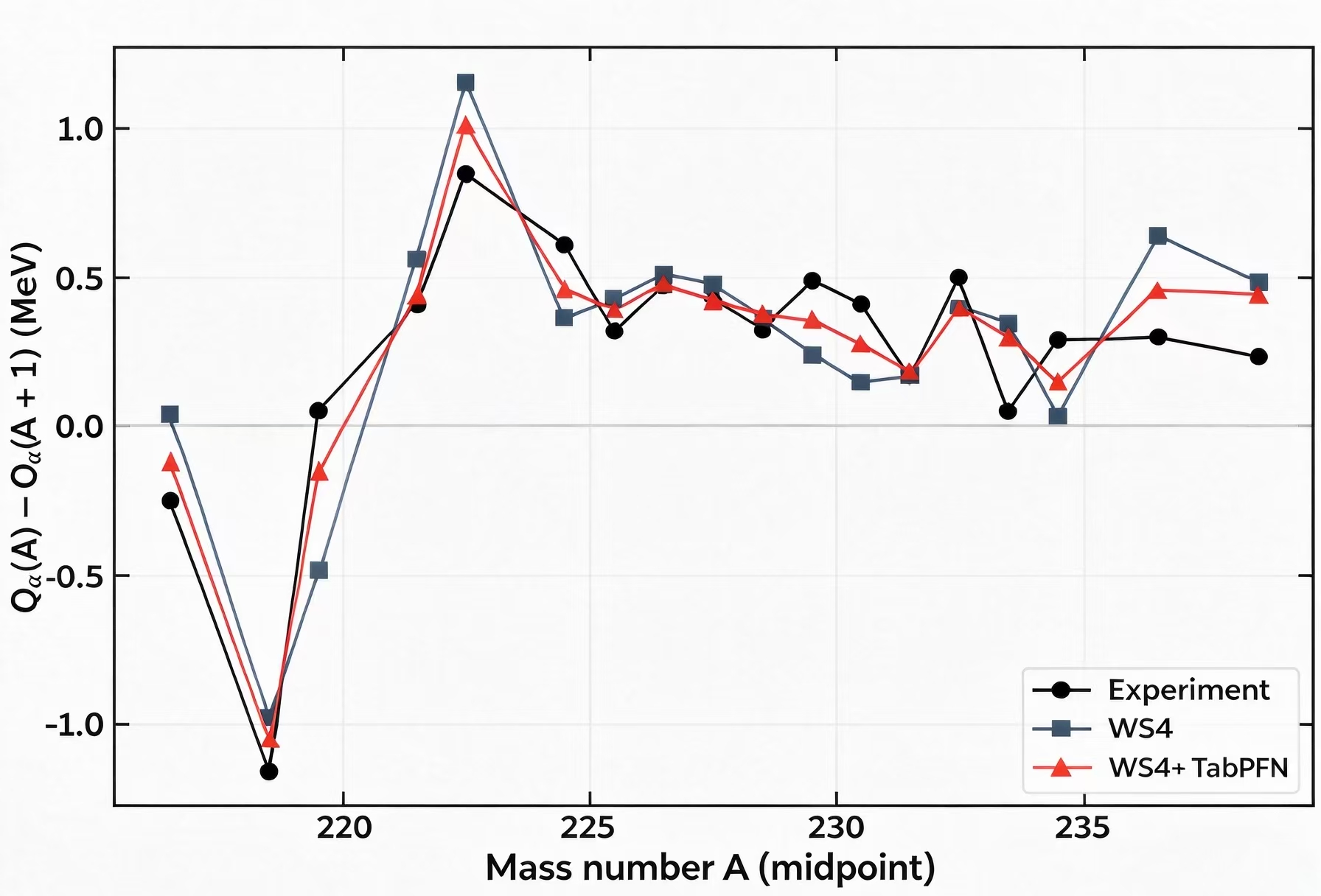}
        \caption{U isotopic chain (\(Z=92\))}
    \end{subfigure}
    
    \caption{\scriptsize Decay energies and odd-even oscillation structures of the three isotopic chains: Po (\(Z = 84\)), Rn (\(Z = 86\)), and U (\(Z = 92\)).}
    \label{fig:iso_chain_qalpha}
\end{figure*}
To check how the correction holds up at different scales, we picked three experimental isotope chains: Po (\(Z = 84\)), Rn (\(Z = 86\)), and U (\(Z = 92\)), all of which cross the \(N = 126\) shell closure. Fig.4(a)-(f) provides two levels of information for each chain pair. The chain plot tracks the evolution of \(Q_\alpha\) with \(A\) to test the overall trend and shell structure. The difference plot shows the oscillation curve of adjacent differences, \(D(A) = Q_\alpha(A) - Q_\alpha(A+1)\). Its amplitude is exactly the odd-even mass staggering (OES) strength, which serves as a check on whether the microscopic structure gets destroyed by the correction.Looking at the overall behavior, \(Q_\alpha\) for all three chains first drops slowly with increasing \(A\), and then exhibits a characteristic peak-valley structure near the \(N = 126\) shell closure. The TabPFN correction does not mess with the shape of the chains or the peak-valley positions at all. It simply shrinks the deviations from experimental values in a systematic way. This basically tells us that the correction acts on the local shell structure rather than the macroscopic trend. On the microscopic side, the oscillation phase after correction matches the experimental data perfectly, and the odd-even zigzag is fully preserved. If the correction were just a pure statistical smoothing, this oscillation would be the first thing to get wiped out. The quantitative comparison in shows Table IV that the amplitude ratios (WS4/exp) for the three chains are 0.83, 0.96, and 1.07, scattered on both sides of 1. After the fix, these numbers tighten to 0.87, 0.96, and 1.02, uniformly approaching the experimental values. This implies that the residual learning just calibrated the amplitude bias of WS4's pairing description, without rewriting its underlying pairing structure.

\begin{equation}
\Delta_{\mathrm{OES}} = \frac{1}{n}\sum_{i=1}^{n} D\left(A_i + \frac{1}{2}\right) = \frac{1}{n}\sum_{i=1}^{n} \left| Q_\alpha(A_i) - Q_\alpha(A_{i+1}) \right|.
\end{equation}

\begin{table}[htbp]
    \centering
    \caption{Average odd-even mass staggering amplitudes for Po, Rn, and U isotopic chains}
    \label{tab:odd_even_staggering}
    \resizebox{\columnwidth}{!}{
        \begin{tabular}{lcccccc}
            \toprule
            Chain & Nuclei & Exp. & WS4 & TabPFN & WS4/Exp. & TabPFN/Exp. \\
            \midrule
            Po (\(Z=84\)) & 33 & 0.3412 & 0.2823 & 0.2982 & 0.827 & 0.874 \\
            Rn (\(Z=86\)) & 33 & 0.3389 & 0.3250 & 0.3252 & 0.958 & 0.959 \\
            U  (\(Z=92\)) & 19 & 0.4063 & 0.4351 & 0.4158 & 1.070 & 1.023 \\
            \bottomrule
        \end{tabular}
    }
\end{table}

Furthermore, using the optimal feature set (g17: A, N, Z, E\(_{\text{def}}\), E\(_{\text{sh}}\), N\(_{\text{p}}\), N\(_{\text{n}}\)) we identified from the ablation study in Section II, we extrapolate the \(\alpha\)-decay energies for nuclides along the Z = 118, 119, and 120 isotopic chains where experimental data are currently missing. The detailed results are listed in Table V and VI.

Based on these decay energy calculations, Fig.5 illustrates how the predicted \(Q_\alpha\) evolves with the neutron number \(N\), comparing the WS4 baseline and the WS4+TabPFN model. As \(N\) increases, we can see a clear pattern. Because the parent nuclei are far from the \(\beta\)-stability line, the symmetry energy term makes a stronger positive contribution, pushing \(Q_\alpha\) upward. On the flip side, a larger mass number enhances Coulomb repulsion. The interplay between these two effects shapes the observed curves: \(Q_\alpha\) first rises and then falls with \(N\), peaking around \(N = 185\text{--}190\). 

Overall, the residual correction from TabPFN does not alter the fundamental shape of these chains. This aligns perfectly with what we discussed in Section III: the ML model inherits the macroscopic trends from WS4 and simply compensates for local microscopic effects, rather than fitting the whole curve from scratch. Furthermore, the overall predicted \(Q_\alpha\) values are consistently higher than the pure WS4 results. From a data-driven standpoint, this systematic positive bias in our predictions is far from random. 

First, the training data for superheavy nuclei with experimental values (Z = 104--117) in the WS4 mass table is quite sparse, so the model has limited material to learn from. Second, within the training set's superheavy region (e.g., Fl, Hs, Ds isotopic chains), the measured \(\alpha\)-decay energies are generally higher than the WS4 predictions. During training, TabPFN successfully picked up on this local systematic error and naturally carried this positive correction trend over when extrapolating to Z = 118--120. Finally, the WS4 macroscopic-microscopic model itself carries uncertainties when calculating microscopic shell corrections in such extreme neutron- or proton-rich regions. By learning the residual distributions from known nuclides, the ML model can effectively compensate for WS4's underestimation of shell corrections or pairing effects in this extreme territory, which likely drives the overall predictions a bit higher.

\begin{figure*}[htbp] 
    \centering
    
    \begin{subfigure}[b]{0.32\textwidth}
        \centering
        \includegraphics[width=\textwidth]{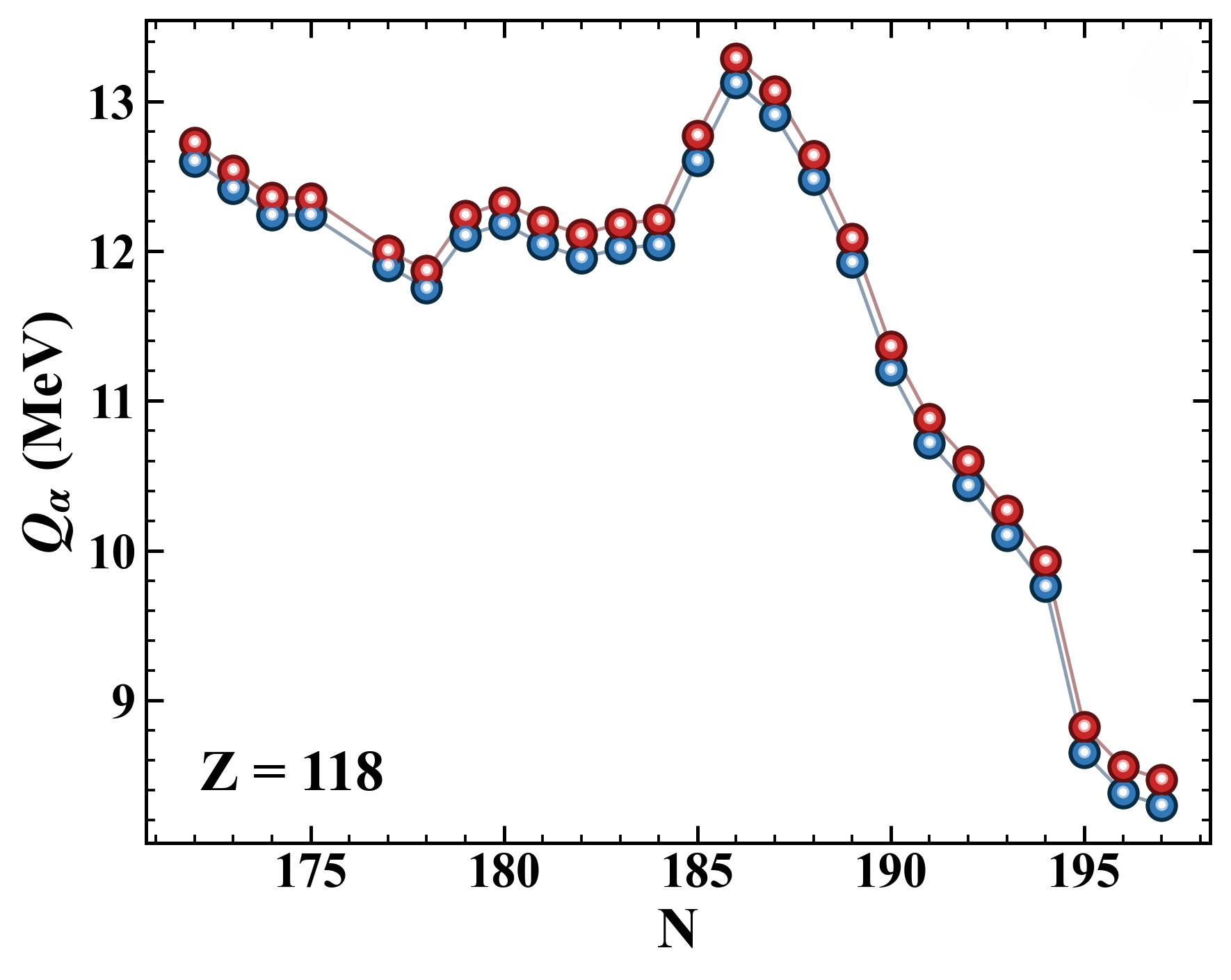} 
        \caption{} 
        \label{fig:Z118}
    \end{subfigure}
    \hfill
    \begin{subfigure}[b]{0.32\textwidth}
        \centering
        \includegraphics[width=\textwidth]{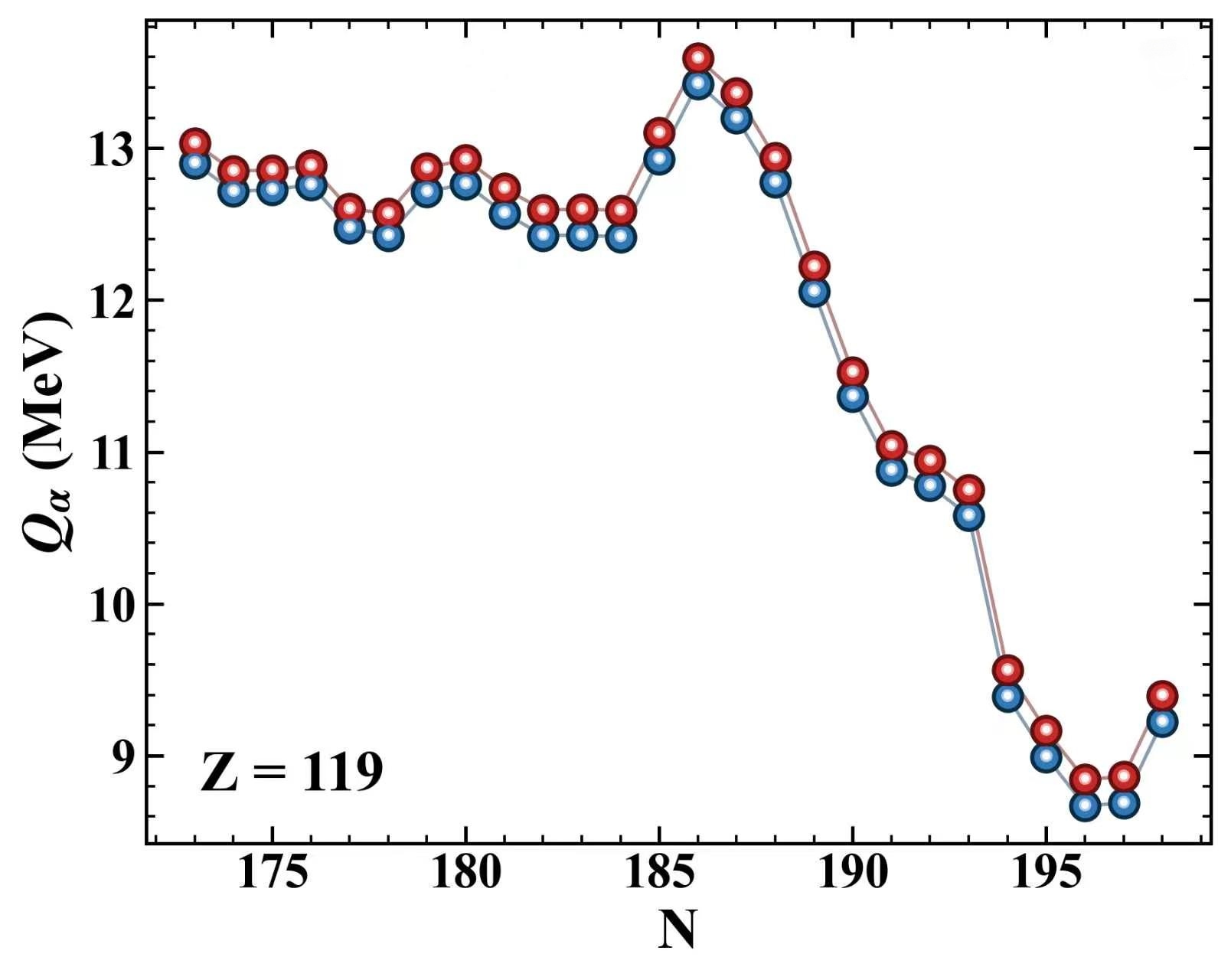} 
        \caption{} 
        \label{fig:Z119}
    \end{subfigure}
    \hfill
    \begin{subfigure}[b]{0.32\textwidth}
        \centering
        \includegraphics[width=\textwidth]{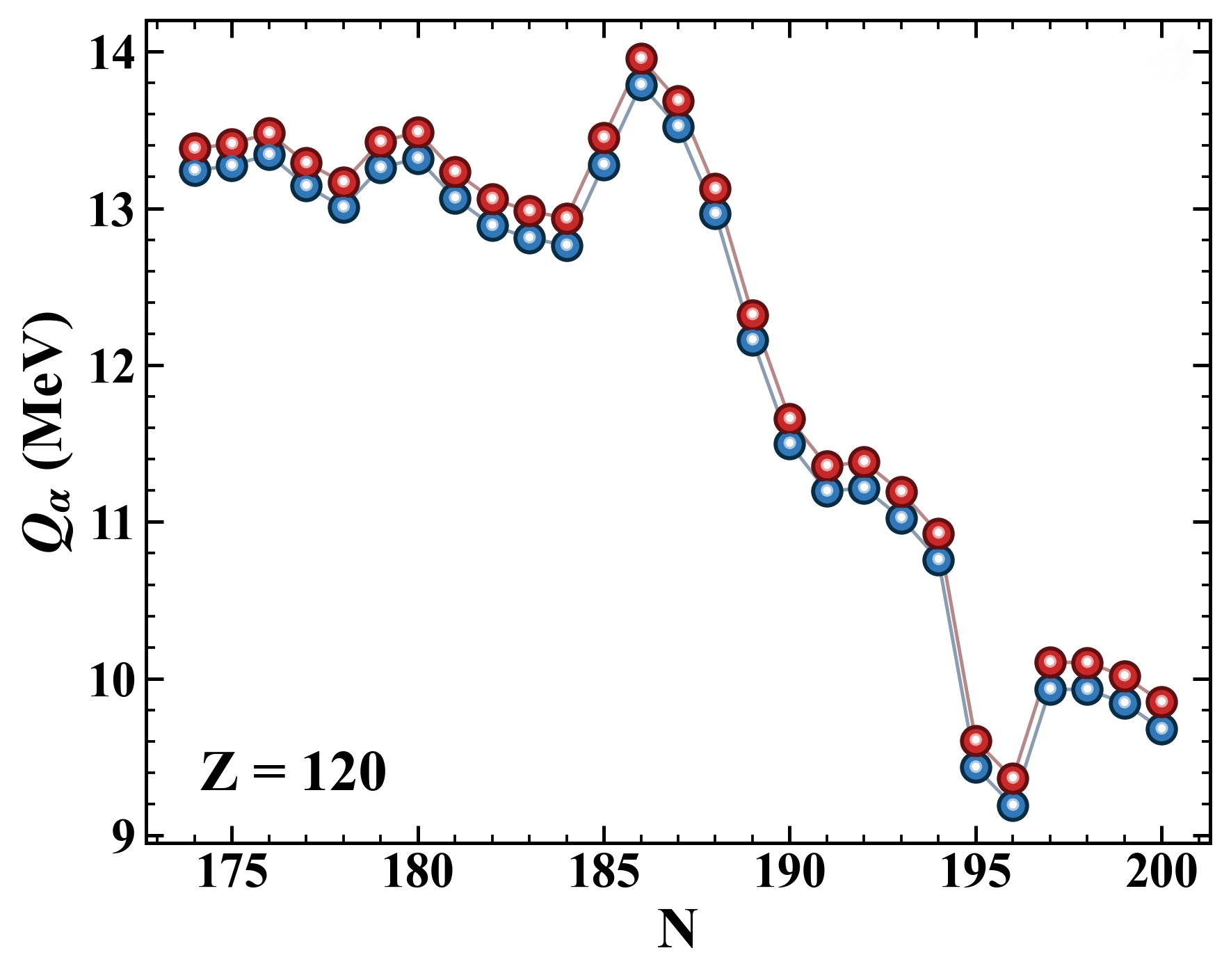} 
        \caption{} 
        \label{fig:Z120}
    \end{subfigure}
    
    \caption{\footnotesize Extrapolation predictions of \(Q_\alpha\) for the Z = 118, 119, and 120 isotopic chains by the WS4 baseline and WS4+TabPFN.}
    \label{fig:extrapolation}
\end{figure*}

\begin{figure*}[htbp] 
    \centering
    \begin{subfigure}[b]{0.32\textwidth}
        \centering
        \includegraphics[width=\textwidth]{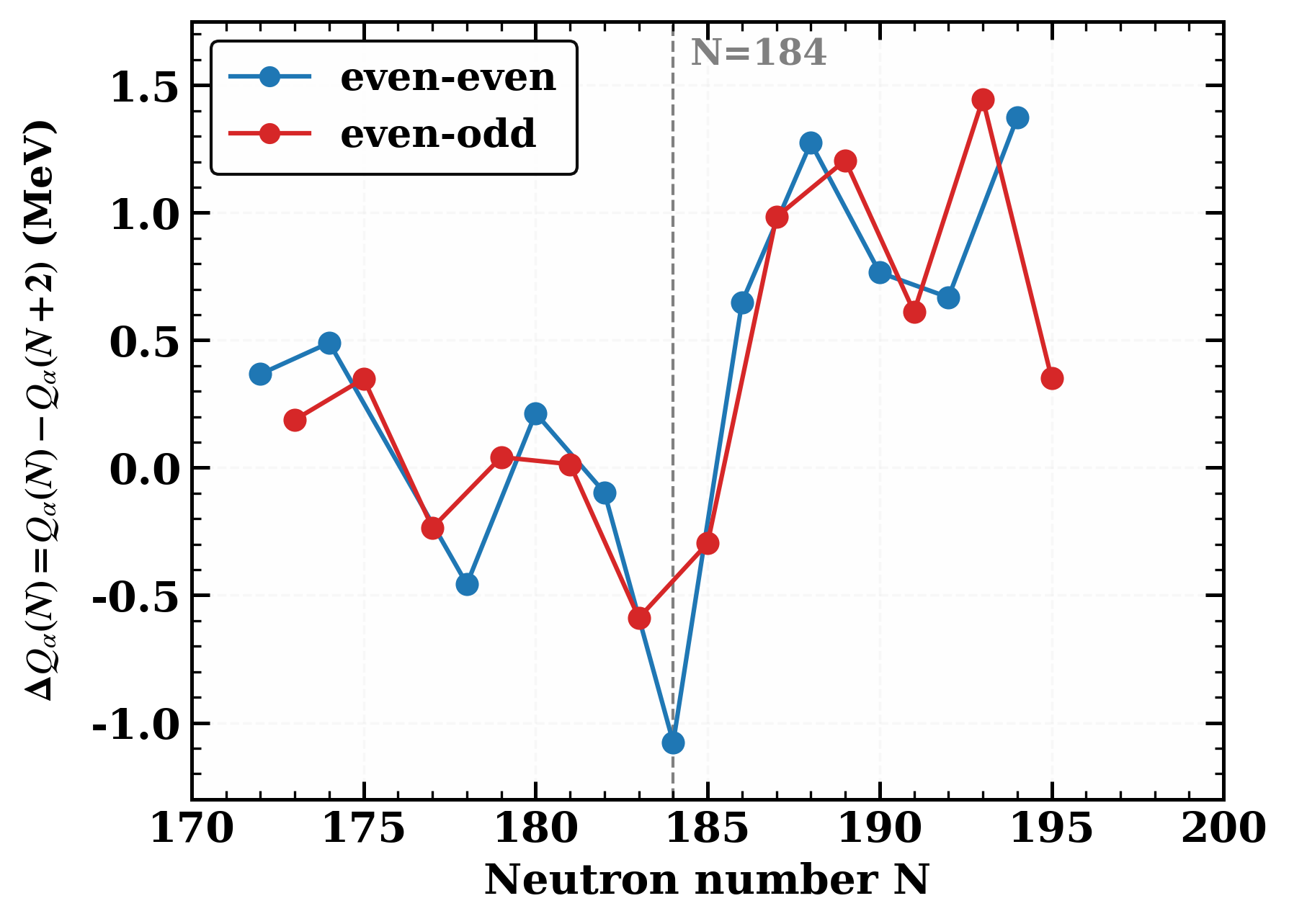} 
        \caption{Z = 118}
        \label{fig:diff_Z118}
    \end{subfigure}
    \hfill
    \begin{subfigure}[b]{0.32\textwidth}
        \centering
        \includegraphics[width=\textwidth]{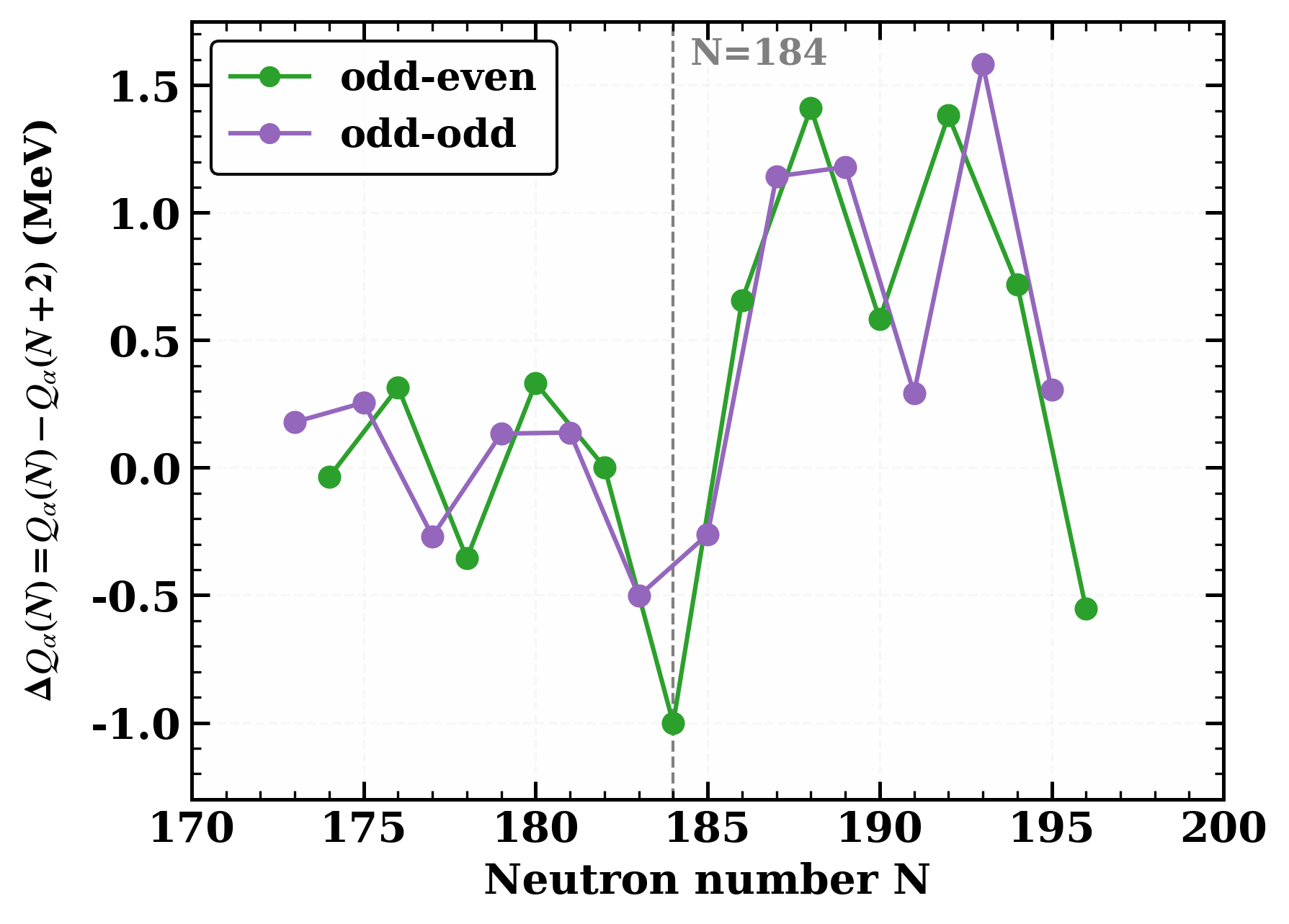} 
        \caption{Z = 119}
        \label{fig:diff_Z119}
    \end{subfigure}
    \hfill
    \begin{subfigure}[b]{0.32\textwidth}
        \centering
        \includegraphics[width=\textwidth]{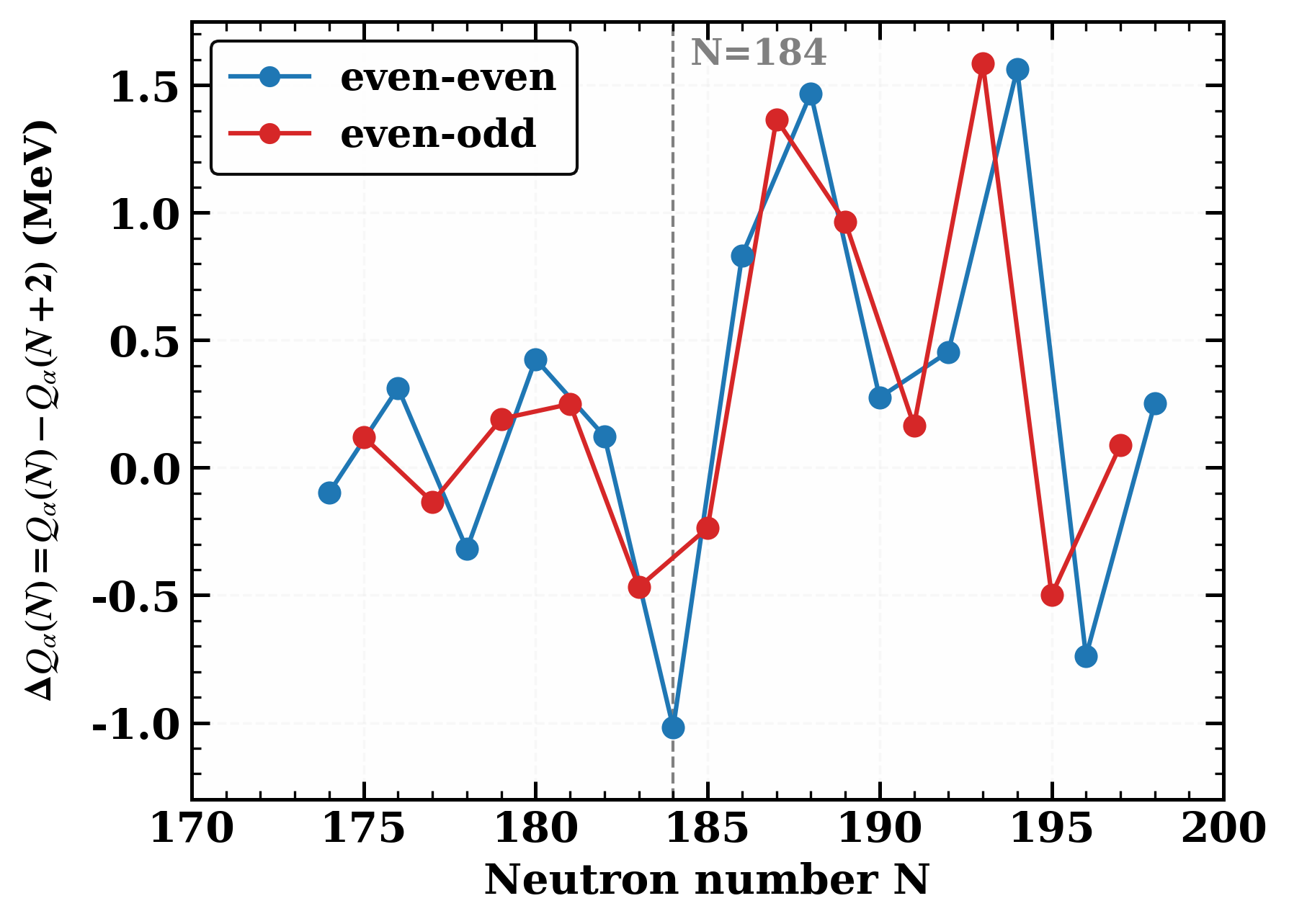} 
        \caption{Z = 120}
        \label{fig:diff_Z120}
    \end{subfigure}
    
    \caption{\footnotesize Difference curves for same-parity nuclides along the three predicted isotopic chains: (a) Z = 118, (b) Z = 119, and (c) Z = 120.}
    \label{fig:all_diff_curves}
\end{figure*}

Beyond the macroscopic trends, we can resolve the local structural features along the chain by taking the difference between adjacent nuclei of the same parity. This is expressed as
\begin{equation}
\Delta Q_\alpha(N) = Q_\alpha(N) - Q_\alpha(N+2).
\end{equation}

In atomic nuclei, the pairing effect of nucleons causes the decay energy to exhibit a strong odd-even staggering with respect to the neutron number \(N\). Because \(N\) and \(N+2\) share the same parity, the pairing energy they contain remains virtually identical. Taking this finite difference basically cancels out the massive pairing energy background, which allows us to cleanly separate the subtle shell effects from the smooth macroscopic trends.

On top of that, odd-\(N\) and even-\(N\) nuclei have a systematic difference in their baseline energy due to unpaired neutrons. That is exactly why we plot the even-\(N\) and odd-\(N\) chains separately here. As Fig.6 shows, the two curves are systematically shifted from each other vertically, yet their oscillation phases are perfectly aligned. This further confirms that our model accurately captures the evolution of the shell structure superimposed on the pairing energy, regardless of whether we are looking at even or odd nuclei.

Nuclear structure theories have long suggested that \(N = 184\) is a neutron magic number in the superheavy region, corresponding to a spherical closed shell. If this holds true, nuclides around \(N = 184\) should exhibit extra shell stabilization, meaning \(Q_\alpha\) would drop to a local minimum compared to their neighbors.

The difference curves for all three chains in this work show a distinct dip right at \(N = 184\). This result perfectly aligns with the theoretical prediction above and matches the original shell correction picture of the WS4 model. It basically tells us that the TabPFN residual correction doesn't mess up the WS4 description of the \(N = 184\) shell effect. On top of that, the curves for even-\(N\) and odd-\(N\) nuclei remain systematically separated, and their oscillation phases are completely preserved.

\begin{table*}[htbp]
    \centering
    \caption{Comparison of theoretical predictions for \(\alpha\)-decay energies of the Z = 118-120 isotopic chains.}
    \label{tab:alpha_decay_1}
    \footnotesize
    \setlength{\tabcolsep}{4pt} 
    
    \begin{tabular*}{\textwidth}{@{\extracolsep{\fill}} ccccccc @{}}
        \toprule
        Transition & E\(_{\text{sh}}\) & E\(_{\text{def}}\) & N\(_{\text{p}}\) & N\(_{\text{n}}\) & Q\(_\alpha\) (WS4) & Q\(_\alpha\) (TabPFN) \\
        \midrule
        2900g & \(-5.6176\) & \(1.6494\) & \(-8\) & 46 & \(12.6007\) & \(12.7265\) \\
        2910g & \(-5.6856\) & \(1.7349\) & \(-8\) & 47 & \(12.4203\) & \(12.5404\) \\
        2920g & \(-6.018\)  & \(1.7449\) & \(-8\) & 48 & \(12.2404\) & \(12.3591\) \\
        2930g & \(-6.014\)  & \(1.5208\) & \(-8\) & 49 & \(12.2419\) & \(12.3523\) \\
        2950g & \(-5.846\)  & \(1.0122\) & \(-8\) & 51 & \(11.9027\) & \(12.0033\) \\
        2960g & \(-5.9272\) & \(0.7642\) & \(-8\) & 52 & \(11.752\)  & \(11.8693\) \\
        2970g & \(-5.5837\) & \(0.5684\) & \(-8\) & 53 & \(12.1038\) & \(12.2388\) \\
        2980g & \(-5.5357\) & \(0.5356\) & \(-8\) & 54 & \(12.1823\) & \(12.3262\) \\
        2990g & \(-5.2979\) & \(0.5378\) & \(-8\) & 55 & \(12.0457\) & \(12.1973\) \\
        3000g & \(-5.2544\) & \(0.4768\) & \(-8\) & 56 & \(11.9556\) & \(12.1123\) \\
        3010g & \(-4.8469\) & \(0.2056\) & \(-8\) & 57 & \(12.0185\) & \(12.1835\) \\
        3020g & \(-4.7807\) & \(0.1108\) & \(-8\) & 58 & \(12.0407\) & \(12.2095\) \\
        3030g & \(-3.7188\) & \(0.1934\) & \(-8\) & 59 & \(12.6048\) & \(12.7725\) \\
        3040g & \(-2.8795\) & \(0.4133\) & \(-8\) & 60 & \(13.1222\) & \(13.2868\) \\
        3050g & \(-2.1064\) & \(0.6746\) & \(-8\) & 61 & \(12.9076\) & \(13.0688\) \\
        3060g & \(-1.5484\) & \(1.0276\) & \(-8\) & 62 & \(12.4799\) & \(12.6375\) \\
        3070g & \(-1.4215\) & \(1.519\)  & \(-8\) & 63 & \(11.9241\) & \(12.0859\) \\
        3080g & \(-1.4121\) & \(2.2578\) & \(-8\) & 64 & \(11.2046\) & \(11.3637\) \\
        3090g & \(-1.0406\) & \(2.7585\) & \(-8\) & 65 & \(10.7194\) & \(10.8806\) \\
        3100g & \(-0.7739\) & \(3.2275\) & \(-8\) & 66 & \(10.4332\) & \(10.5984\) \\
        3110g & \(-0.7412\) & \(3.7009\) & \(-8\) & 67 & \(10.1019\) & \(10.2699\) \\
        3120g & \(-1.5037\) & \(4.3013\) & \(-8\) & 68 & \(9.7608\)  & \(9.9303\) \\
        3130g & \(-2.2694\) & \(5.7227\) & \(-8\) & 69 & \(8.6505\)  & \(8.8248\) \\
        3140g & \(-2.3498\) & \(6.2959\) & \(-8\) & 70 & \(8.3834\)  & \(8.5577\) \\
        3150g & \(-2.1417\) & \(6.7761\) & \(-8\) & 71 & \(8.2994\)  & \(8.4735\) \\
        292Uue & \(-5.7385\) & \(1.6716\) & \(-7\) & 47 & \(12.9016\) & \(13.035\) \\
        293Uue & \(-6.0344\) & \(1.6997\) & \(-7\) & 48 & \(12.7152\) & \(12.8489\) \\
        294Uue & \(-6.0757\) & \(1.4731\) & \(-7\) & 49 & \(12.7261\) & \(12.856\) \\
        295Uue & \(-6.118\)  & \(1.1401\) & \(-7\) & 50 & \(12.7585\) & \(12.8846\) \\
        296Uue & \(-5.7891\) & \(0.8796\) & \(-7\) & 51 & \(12.4752\) & \(12.6008\) \\
        297Uue & \(-5.9313\) & \(0.6355\) & \(-7\) & 52 & \(12.4242\) & \(12.5694\) \\
        298Uue & \(-5.4633\) & \(0.3916\) & \(-7\) & 53 & \(12.7134\) & \(12.8695\) \\
        299Uue & \(-5.4493\) & \(0.353\)  & \(-7\) & 54 & \(12.7643\) & \(12.9247\) \\
        300Uue & \(-5.1698\) & \(0.3601\) & \(-7\) & 55 & \(12.5719\) & \(12.7358\) \\
        301Uue & \(-5.1576\) & \(0.3252\) & \(-7\) & 56 & \(12.4257\) & \(12.5927\) \\
        302Uue & \(-4.7998\) & \(0.1008\) & \(-7\) & 57 & \(12.4259\) & \(12.5979\) \\
        303Uue & \(-4.7705\) & \(0.0356\) & \(-7\) & 58 & \(12.4161\) & \(12.5903\) \\
        304Uue & \(-3.6947\) & \(0.1172\) & \(-7\) & 59 & \(12.9289\) & \(13.1005\) \\
        305Uue & \(-2.9692\) & \(0.3508\) & \(-7\) & 60 & \(13.4245\) & \(13.5925\) \\
        306Uue & \(-2.1347\) & \(0.615\)  & \(-7\) & 61 & \(13.1984\) & \(13.3623\) \\
        307Uue & \(-1.618\)  & \(0.9749\) & \(-7\) & 62 & \(12.778\)  & \(12.9361\) \\
        308Uue & \(-1.5956\) & \(1.6599\) & \(-7\) & 63 & \(12.0574\) & \(12.2202\) \\
        309Uue & \(-1.5618\) & \(2.3955\) & \(-7\) & 64 & \(11.3666\) & \(11.5259\) \\
        310Uue & \(-1.2705\) & \(2.9202\) & \(-7\) & 65 & \(10.879\)  & \(11.041\) \\
        311Uue & \(-1.0296\) & \(3.3784\) & \(-7\) & 66 & \(10.7769\) & \(10.9432\) \\
        312Uue & \(-0.7996\) & \(3.7983\) & \(-7\) & 67 & \(10.5806\) & \(10.7486\) \\
        313Uue & \(-2.5109\) & \(5.1992\) & \(-7\) & 68 & \(9.389\)   & \(9.5621\) \\
        314Uue & \(-2.5222\) & \(5.7834\) & \(-7\) & 69 & \(8.9928\)  & \(9.1668\) \\
        315Uue & \(-2.6122\) & \(6.3799\) & \(-7\) & 70 & \(8.6697\)  & \(8.8436\) \\
        316Uue & \(-2.3602\) & \(6.9272\) & \(-7\) & 71 & \(8.6878\)  & \(8.8615\) \\
        317Uue & \(-2.2216\) & \(7.3643\) & \(-7\) & 72 & \(9.2224\)  & \(9.3958\) \\
        \bottomrule
    \end{tabular*}
\end{table*}

\begin{table*}[htbp]
    \centering
    \caption{Comparison of theoretical predictions for \(\alpha\)-decay energies of the Z = 118-120 isotopic chains.}
    \label{tab:alpha_decay_2}
    \footnotesize
    \setlength{\tabcolsep}{4pt} 
    
    \begin{tabular*}{\textwidth}{@{\extracolsep{\fill}} ccccccc @{}}
        \toprule
        Transition & E\(_{\text{sh}}\) & E\(_{\text{def}}\) & N\(_{\text{p}}\) & N\(_{\text{n}}\) & Q\(_\alpha\) (WS4) & Q\(_\alpha\) (TabPFN) \\
        \midrule
        294Ubn & \(-6.1601\) & \(1.5808\) & \(-6\) & 48 & \(13.2418\) & \(13.3826\) \\
        295Ubn & \(-6.1639\) & \(1.3599\) & \(-6\) & 49 & \(13.2718\) & \(13.4108\) \\
        296Ubn & \(-6.2324\) & \(1.0229\) & \(-6\) & 50 & \(13.343\)  & \(13.4798\) \\
        297Ubn & \(-5.785\)  & \(0.7122\) & \(-6\) & 51 & \(13.1433\) & \(13.2897\) \\
        298Ubn & \(-5.8879\) & \(0.4539\) & \(-6\) & 52 & \(13.0069\) & \(13.1669\) \\
        299Ubn & \(-5.4787\) & \(0.2388\) & \(-6\) & 53 & \(13.2593\) & \(13.4248\) \\
        300Ubn & \(-5.2681\) & \(0.1777\) & \(-6\) & 54 & \(13.3186\) & \(13.4861\) \\
        301Ubn & \(-5.1413\) & \(0.2124\) & \(-6\) & 55 & \(13.0648\) & \(13.2332\) \\
        302Ubn & \(-5.1578\) & \(0.2024\) & \(-6\) & 56 & \(12.8899\) & \(13.0604\) \\
        303Ubn & \(-4.8818\) & \(0.0439\) & \(-6\) & 57 & \(12.8102\) & \(12.9833\) \\
        304Ubn & \(-4.9018\) & \(0.0078\) & \(-6\) & 58 & \(12.763\)  & \(12.9372\) \\
        305Ubn & \(-3.8165\) & \(0.0847\) & \(-6\) & 59 & \(13.279\)  & \(13.4516\) \\
        306Ubn & \(-3.0433\) & \(0.3093\) & \(-6\) & 60 & \(13.7874\) & \(13.9562\) \\
        307Ubn & \(-2.3065\) & \(0.5789\) & \(-6\) & 61 & \(13.5231\) & \(13.6876\) \\
        308Ubn & \(-1.9823\) & \(1.0757\) & \(-6\) & 62 & \(12.9661\) & \(13.1246\) \\
        309Ubn & \(-1.8854\) & \(1.8572\) & \(-6\) & 63 & \(12.1601\) & \(12.3212\) \\
        310Ubn & \(-1.8928\) & \(2.5889\) & \(-6\) & 64 & \(11.4988\) & \(11.6572\) \\
        311Ubn & \(-1.5853\) & \(3.1042\) & \(-6\) & 65 & \(11.1955\) & \(11.3574\) \\
        312Ubn & \(-1.2562\) & \(3.5388\) & \(-6\) & 66 & \(11.2162\) & \(11.3817\) \\
        313Ubn & \(-0.9722\) & \(3.9592\) & \(-6\) & 67 & \(11.0248\) & \(11.1922\) \\
        314Ubn & \(-0.8247\) & \(4.4189\) & \(-6\) & 68 & \(10.7586\) & \(10.9275\) \\
        315Ubn & \(-2.8735\) & \(5.9385\) & \(-6\) & 69 & \(9.4336\)  & \(9.6061\) \\
        316Ubn & \(-2.9909\) & \(6.4944\) & \(-6\) & 70 & \(9.1917\)  & \(9.3642\) \\
        317Ubn & \(-2.7614\) & \(6.9253\) & \(-6\) & 71 & \(9.933\)   & \(10.1054\) \\
        318Ubn & \(-2.6024\) & \(7.3296\) & \(-6\) & 72 & \(9.9308\)  & \(10.1031\) \\
        319Ubn & \(-2.3881\) & \(7.6846\) & \(-6\) & 73 & \(9.845\)   & \(10.0173\) \\
        320Ubn & \(-2.2813\) & \(7.97\)   & \(-6\) & 74 & \(9.6782\)  & \(9.8509\) \\
        \bottomrule
    \end{tabular*}
\end{table*}

\section{\quad Summary}
In this work, we developed a physics-guided residual learning framework combining the macroscopic–microscopic mass model WS4 with the Transformer-based tabular foundation model TabPFN to improve the prediction of \(\alpha\)-decay energies (\(Q_\alpha\)) in heavy and superheavy nuclei. By learning the discrepancies between WS4 predictions and experimental values, the framework improves predictive accuracy while retaining the underlying nuclear-structure trends. Through 18 physics-informed feature ablation experiments, the optimal feature set g17, comprising \((A,N,Z,E_{\mathrm{def}},E_{\mathrm{sh}},N_p,N_n)\), was identified. Deformation energy emerged as the most influential individual feature, while substantial redundancy was observed among shell-related descriptors. With g17, the framework achieved RMSEs of 0.1379, 0.1689, and 0.1442 MeV for the training, validation, and combined datasets, respectively, reducing the overall RMSE by 47.43$\%$ relative to WS4. Notably, using only \((A,N,Z)\) reduced the RMSE to 0.1588 MeV, indicating that much of the WS4 residual follows systematic trends in nucleon numbers, while deformation- and shell-related features provide further corrections to local nuclear-structure effects.

The framework consistently improved predictions across all four proton–neutron odd–even configurations, with particularly notable gains for odd-mass nuclei. Analyses of the Po, Rn, and U isotopic chains near the \(N=126\) shell closure demonstrated improved agreement with experimental \(Q_\alpha\) values while preserving shell-related variations and odd–even staggering patterns. Extrapolation to superheavy nuclei with \(Z=118\text{--}120\) revealed physically consistent trends, with \(Q_\alpha\) reaching a maximum around \(N\approx185\text{--}190\). The characteristic minimum in same-parity \(Q_\alpha\) differences near \(N=184\) was also retained, indicating that the correction preserves the predicted neutron shell effect. Although the corrected values are systematically higher than WS4 predictions in this region, the shift is consistent with residual trends observed in experimentally known superheavy nuclei. Overall, the WS4+TabPFN framework substantially improves predictive accuracy while maintaining essential nuclear-structure features, offering a promising approach for estimating \(\alpha\)-decay energies and supporting half-life predictions in experimentally unexplored superheavy nuclei.



\begin{center}
    \textbf{ACKNOWLEDGMENTS}
\end{center}

This work is supported by Yunnan Fundamental Research Projects (No. 202501AT070067, 202401AU070074 and 202501CF070189), Yunnan Provincial Xing Dian Talent Support Program (Young Talents Special Program, (Young Talents Special Program, No. XDYC-QNRC-2023-0162), Kunming University Talent Introduction Research Project (No. YJL24019), Yunnan Provincial Department of Education Scientific Research Fund Project (No. 2025Y1055 and 2025Y1042), the Special Basic Cooperative Research Programs of Yunnan ProvincialUndergraduate Universities’ Association (NO. 202101BA070001-144), the Program for Frontier Research Team of Kunming University 2023, National Natural Science Foundation of China (No. 12063006), National College Student Innovation and Entrepreneurship Training Program (No. 202511393011, 202511393012, and 202511393015), Yunnan Province College Student Innovation and Entrepreneurship Training Program (No. S202511393003, S202511393043, and S202511393044), and Xing Dingyu Academician Workstation of Yunnan Province (No. 202605AF350035).

\bibliography{Reference}

\begin{thebibliography}{60}%
\makeatletter
\providecommand \@ifxundefined [1]{%
 \@ifx{#1\undefined}
}%
\providecommand \@ifnum [1]{%
 \ifnum #1\expandafter \@firstoftwo
 \else \expandafter \@secondoftwo
 \fi
}%
\providecommand \@ifx [1]{%
 \ifx #1\expandafter \@firstoftwo
 \else \expandafter \@secondoftwo
 \fi
}%
\providecommand \natexlab [1]{#1}%
\providecommand \enquote  [1]{``#1''}%
\providecommand \bibnamefont  [1]{#1}%
\providecommand \bibfnamefont [1]{#1}%
\providecommand \citenamefont [1]{#1}%
\providecommand \href@noop [0]{\@secondoftwo}%
\providecommand \href [0]{\begingroup \@sanitize@url \@href}%
\providecommand \@href[1]{\@@startlink{#1}\@@href}%
\providecommand \@@href[1]{\endgroup#1\@@endlink}%
\providecommand \@sanitize@url [0]{\catcode `\\12\catcode `\$12\catcode `\&12\catcode `\#12\catcode `\^12\catcode `\_12\catcode `\%12\relax}%
\providecommand \@@startlink[1]{}%
\providecommand \@@endlink[0]{}%
\providecommand \url  [0]{\begingroup\@sanitize@url \@url }%
\providecommand \@url [1]{\endgroup\@href {#1}{\urlprefix }}%
\providecommand \urlprefix  [0]{URL }%
\providecommand \Eprint [0]{\href }%
\providecommand \doibase [0]{https://doi.org/}%
\providecommand \selectlanguage [0]{\@gobble}%
\providecommand \bibinfo  [0]{\@secondoftwo}%
\providecommand \bibfield  [0]{\@secondoftwo}%
\providecommand \translation [1]{[#1]}%
\providecommand \BibitemOpen [0]{}%
\providecommand \bibitemStop [0]{}%
\providecommand \bibitemNoStop [0]{.\EOS\space}%
\providecommand \EOS [0]{\spacefactor3000\relax}%
\providecommand \BibitemShut  [1]{\csname bibitem#1\endcsname}%
\let\auto@bib@innerbib\@empty
\bibitem [{\citenamefont {Rutherford}\ and\ \citenamefont {Soddy}(1902)}]{1}%
  \BibitemOpen
  \bibfield  {author} {\bibinfo {author} {\bibfnamefont {E.}~\bibnamefont {Rutherford}}\ and\ \bibinfo {author} {\bibfnamefont {F.}~\bibnamefont {Soddy}},\ }\bibfield  {title} {\bibinfo {title} {{The} cause and nature of radioactivity},\ }\href@noop {} {\bibfield  {journal} {\bibinfo  {journal} {The London, Edinburgh, and Dublin Philosophical Magazine and Journal of Science}\ }\textbf {\bibinfo {volume} {4}},\ \bibinfo {pages} {370} (\bibinfo {year} {1902})}\BibitemShut {NoStop}%
\bibitem [{\citenamefont {Rutherford}\ and\ \citenamefont {Royds}(1909)}]{2}%
  \BibitemOpen
  \bibfield  {author} {\bibinfo {author} {\bibfnamefont {E.}~\bibnamefont {Rutherford}}\ and\ \bibinfo {author} {\bibfnamefont {T.}~\bibnamefont {Royds}},\ }\bibfield  {title} {\bibinfo {title} {{The} nature of the {$\alpha$} particle from radioactive substances},\ }\href@noop {} {\bibfield  {journal} {\bibinfo  {journal} {The London, Edinburgh, and Dublin Philosophical Magazine and Journal of Science}\ }\textbf {\bibinfo {volume} {17}},\ \bibinfo {pages} {281} (\bibinfo {year} {1909})}\BibitemShut {NoStop}%
\bibitem [{\citenamefont {Geiger}\ and\ \citenamefont {Nuttall}(1911)}]{3}%
  \BibitemOpen
  \bibfield  {author} {\bibinfo {author} {\bibfnamefont {H.}~\bibnamefont {Geiger}}\ and\ \bibinfo {author} {\bibfnamefont {J.}~\bibnamefont {Nuttall}},\ }\bibfield  {title} {\bibinfo {title} {{The} ranges of the {$\alpha$} particles from various radioactive substances and a relation between range and period of transformation},\ }\href@noop {} {\bibfield  {journal} {\bibinfo  {journal} {The London, Edinburgh, and Dublin Philosophical Magazine and Journal of Science}\ }\textbf {\bibinfo {volume} {22}},\ \bibinfo {pages} {613} (\bibinfo {year} {1911})}\BibitemShut {NoStop}%
\bibitem [{\citenamefont {Delion}(2009)}]{4}%
  \BibitemOpen
  \bibfield  {author} {\bibinfo {author} {\bibfnamefont {D.~S.}\ \bibnamefont {Delion}},\ }\bibfield  {title} {\bibinfo {title} {Universal decay rule for reduced widths},\ }\href@noop {} {\bibfield  {journal} {\bibinfo  {journal} {Physical Review C}\ }\textbf {\bibinfo {volume} {80}},\ \bibinfo {pages} {024310} (\bibinfo {year} {2009})}\BibitemShut {NoStop}%
\bibitem [{\citenamefont {Oganessian}\ and\ \citenamefont {Utyonkov}(2015)}]{5}%
  \BibitemOpen
  \bibfield  {author} {\bibinfo {author} {\bibfnamefont {Y.~T.}\ \bibnamefont {Oganessian}}\ and\ \bibinfo {author} {\bibfnamefont {V.~K.}\ \bibnamefont {Utyonkov}},\ }\bibfield  {title} {\bibinfo {title} {Superheavy nuclei from {$^{48}\mathrm{Ca}$}-induced reactions},\ }\href {https://doi.org/10.1016/j.nuclphysa.2015.06.002} {\bibfield  {journal} {\bibinfo  {journal} {Nuclear Physics A}\ }\textbf {\bibinfo {volume} {944}},\ \bibinfo {pages} {62} (\bibinfo {year} {2015})}\BibitemShut {NoStop}%
\bibitem [{\citenamefont {Delion}\ \emph {et~al.}(2014)\citenamefont {Delion}, \citenamefont {Liotta}, \citenamefont {Qi},\ and\ \citenamefont {Wyss}}]{6}%
  \BibitemOpen
  \bibfield  {author} {\bibinfo {author} {\bibfnamefont {D.~S.}\ \bibnamefont {Delion}}, \bibinfo {author} {\bibfnamefont {R.~J.}\ \bibnamefont {Liotta}}, \bibinfo {author} {\bibfnamefont {C.}~\bibnamefont {Qi}},\ and\ \bibinfo {author} {\bibfnamefont {R.}~\bibnamefont {Wyss}},\ }\bibfield  {title} {\bibinfo {title} {Probing shape coexistence by {$\alpha$} decays to {$0^{+}$} states},\ }\href {https://doi.org/10.1103/PhysRevC.90.061303} {\bibfield  {journal} {\bibinfo  {journal} {Physical Review C}\ }\textbf {\bibinfo {volume} {90}},\ \bibinfo {pages} {061303} (\bibinfo {year} {2014})}\BibitemShut {NoStop}%
\bibitem [{\citenamefont {Karol}\ \emph {et~al.}(2016)\citenamefont {Karol}, \citenamefont {Barber}, \citenamefont {Sherrill}, \citenamefont {Vardaci},\ and\ \citenamefont {Yamazaki}}]{7}%
  \BibitemOpen
  \bibfield  {author} {\bibinfo {author} {\bibfnamefont {P.~J.}\ \bibnamefont {Karol}}, \bibinfo {author} {\bibfnamefont {R.~C.}\ \bibnamefont {Barber}}, \bibinfo {author} {\bibfnamefont {B.~M.}\ \bibnamefont {Sherrill}}, \bibinfo {author} {\bibfnamefont {E.}~\bibnamefont {Vardaci}},\ and\ \bibinfo {author} {\bibfnamefont {T.}~\bibnamefont {Yamazaki}},\ }\bibfield  {title} {\bibinfo {title} {Discovery of the elements with atomic numbers {{Z}} = 113, 115 and 117 ({{IUPAC Technical Report}})},\ }\href@noop {} {\bibfield  {journal} {\bibinfo  {journal} {Pure and Applied Chemistry}\ }\textbf {\bibinfo {volume} {88}},\ \bibinfo {pages} {139} (\bibinfo {year} {2016})}\BibitemShut {NoStop}%
\bibitem [{\citenamefont {Oganessian}\ \emph {et~al.}(2010)\citenamefont {Oganessian}, \citenamefont {Abdullin} \emph {et~al.}}]{8}%
  \BibitemOpen
  \bibfield  {author} {\bibinfo {author} {\bibfnamefont {Y.~T.}\ \bibnamefont {Oganessian}}, \bibinfo {author} {\bibfnamefont {F.~S.}\ \bibnamefont {Abdullin}}, \emph {et~al.},\ }\bibfield  {title} {\bibinfo {title} {Synthesis of a new element with atomic number {$Z=117$}},\ }\href {https://doi.org/10.1103/PhysRevLett.104.142502} {\bibfield  {journal} {\bibinfo  {journal} {Physical Review Letters}\ }\textbf {\bibinfo {volume} {104}},\ \bibinfo {pages} {142502} (\bibinfo {year} {2010})}\BibitemShut {NoStop}%
\bibitem [{\citenamefont {Ni}\ and\ \citenamefont {Ren}(2012)}]{9}%
  \BibitemOpen
  \bibfield  {author} {\bibinfo {author} {\bibfnamefont {D.}~\bibnamefont {Ni}}\ and\ \bibinfo {author} {\bibfnamefont {Z.}~\bibnamefont {Ren}},\ }\bibfield  {title} {\bibinfo {title} {Binding energies, {$\alpha$}-decay energies, and {$\alpha$}-decay half-lives for heavy and superheavy nuclei},\ }\href@noop {} {\bibfield  {journal} {\bibinfo  {journal} {Nuclear Physics A}\ }\textbf {\bibinfo {volume} {893}},\ \bibinfo {pages} {13} (\bibinfo {year} {2012})}\BibitemShut {NoStop}%
\bibitem [{\citenamefont {Wang}\ \emph {et~al.}(2015)\citenamefont {Wang}, \citenamefont {Wang}, \citenamefont {Hou},\ and\ \citenamefont {Gu}}]{10}%
  \BibitemOpen
  \bibfield  {author} {\bibinfo {author} {\bibfnamefont {Y.~Z.}\ \bibnamefont {Wang}}, \bibinfo {author} {\bibfnamefont {S.~J.}\ \bibnamefont {Wang}}, \bibinfo {author} {\bibfnamefont {Z.~Y.}\ \bibnamefont {Hou}},\ and\ \bibinfo {author} {\bibfnamefont {J.~Z.}\ \bibnamefont {Gu}},\ }\bibfield  {title} {\bibinfo {title} {Systematic study of {$\alpha$} -decay energies and half-lives of superheavy nuclei},\ }\href@noop {} {\bibfield  {journal} {\bibinfo  {journal} {Physical Review C}\ }\textbf {\bibinfo {volume} {92}},\ \bibinfo {pages} {064301} (\bibinfo {year} {2015})}\BibitemShut {NoStop}%
\bibitem [{\citenamefont {Qi}\ \emph {et~al.}(2025)\citenamefont {Qi}, \citenamefont {Wang}, \citenamefont {Cui}, \citenamefont {Wang},\ and\ \citenamefont {Gu}}]{11}%
  \BibitemOpen
  \bibfield  {author} {\bibinfo {author} {\bibfnamefont {L.-Q.}\ \bibnamefont {Qi}}, \bibinfo {author} {\bibfnamefont {H.-M.}\ \bibnamefont {Wang}}, \bibinfo {author} {\bibfnamefont {J.-P.}\ \bibnamefont {Cui}}, \bibinfo {author} {\bibfnamefont {Y.-Z.}\ \bibnamefont {Wang}},\ and\ \bibinfo {author} {\bibfnamefont {J.-Z.}\ \bibnamefont {Gu}},\ }\bibfield  {title} {\bibinfo {title} {{\emph{{$\alpha$}}} -decay half-lives of superheavy nuclei within a one-parameter model},\ }\href@noop {} {\bibfield  {journal} {\bibinfo  {journal} {Chinese Physics C}\ }\textbf {\bibinfo {volume} {49}},\ \bibinfo {pages} {094102} (\bibinfo {year} {2025})}\BibitemShut {NoStop}%
\bibitem [{\citenamefont {Hu}\ and\ \citenamefont {Wu}(2025)}]{12}%
  \BibitemOpen
  \bibfield  {author} {\bibinfo {author} {\bibfnamefont {J.}~\bibnamefont {Hu}}\ and\ \bibinfo {author} {\bibfnamefont {C.}~\bibnamefont {Wu}},\ }\bibfield  {title} {\bibinfo {title} {Nonlocality effect in {$\alpha$} decay half-lives for even-even nuclei within a two potential approach},\ }\href@noop {} {\bibfield  {journal} {\bibinfo  {journal} {The European Physical Journal A}\ }\textbf {\bibinfo {volume} {61}},\ \bibinfo {pages} {129} (\bibinfo {year} {2025})}\BibitemShut {NoStop}%
\bibitem [{\citenamefont {{Rui-Wang}}\ \emph {et~al.}(2017)\citenamefont {{Rui-Wang}}, \citenamefont {Wang}, \citenamefont {Qian},\ and\ \citenamefont {Ren}}]{13}%
  \BibitemOpen
  \bibfield  {author} {\bibinfo {author} {\bibnamefont {{Rui-Wang}}}, \bibinfo {author} {\bibfnamefont {R.-Y.}\ \bibnamefont {Wang}}, \bibinfo {author} {\bibfnamefont {Y.-B.}\ \bibnamefont {Qian}},\ and\ \bibinfo {author} {\bibfnamefont {Z.-Z.}\ \bibnamefont {Ren}},\ }\bibfield  {title} {\bibinfo {title} {Signatures of shell evolution in alpha decay across the {{N}} = 126 shell closure},\ }\href@noop {} {\bibfield  {journal} {\bibinfo  {journal} {Chinese Physics C}\ }\textbf {\bibinfo {volume} {41}},\ \bibinfo {pages} {064103} (\bibinfo {year} {2017})}\BibitemShut {NoStop}%
\bibitem [{\citenamefont {Deng}\ and\ \citenamefont {Zhang}(2021)}]{14}%
  \BibitemOpen
  \bibfield  {author} {\bibinfo {author} {\bibfnamefont {J.-G.}\ \bibnamefont {Deng}}\ and\ \bibinfo {author} {\bibfnamefont {H.-F.}\ \bibnamefont {Zhang}},\ }\bibfield  {title} {\bibinfo {title} {Correlation between {$\alpha$}-particle preformation factor and {$\alpha$} decay energy},\ }\href@noop {} {\bibfield  {journal} {\bibinfo  {journal} {Physics Letters B}\ }\textbf {\bibinfo {volume} {816}},\ \bibinfo {pages} {136247} (\bibinfo {year} {2021})}\BibitemShut {NoStop}%
\bibitem [{\citenamefont {M{\"o}ller}\ \emph {et~al.}(1995)\citenamefont {M{\"o}ller}, \citenamefont {Nix}, \citenamefont {Myers},\ and\ \citenamefont {Swiatecki}}]{15}%
  \BibitemOpen
  \bibfield  {author} {\bibinfo {author} {\bibfnamefont {P.}~\bibnamefont {M{\"o}ller}}, \bibinfo {author} {\bibfnamefont {J.~R.}\ \bibnamefont {Nix}}, \bibinfo {author} {\bibfnamefont {W.~D.}\ \bibnamefont {Myers}},\ and\ \bibinfo {author} {\bibfnamefont {W.~J.}\ \bibnamefont {Swiatecki}},\ }\bibfield  {title} {\bibinfo {title} {Nuclear {{Ground-State Masses}} and {{Deformations}}},\ }\href@noop {} {\bibfield  {journal} {\bibinfo  {journal} {Atomic Data and Nuclear Data Tables}\ }\textbf {\bibinfo {volume} {59}},\ \bibinfo {pages} {185} (\bibinfo {year} {1995})}\BibitemShut {NoStop}%
\bibitem [{\citenamefont {Olsen}\ and\ \citenamefont {Nazarewicz}(2019)}]{16}%
  \BibitemOpen
  \bibfield  {author} {\bibinfo {author} {\bibfnamefont {E.}~\bibnamefont {Olsen}}\ and\ \bibinfo {author} {\bibfnamefont {W.}~\bibnamefont {Nazarewicz}},\ }\bibfield  {title} {\bibinfo {title} {Alpha {{Decay Energies}} of {{Superheavy Nuclei}}: {{Systematic Trends}}},\ }\href@noop {} {\bibfield  {journal} {\bibinfo  {journal} {Physical Review C}\ }\textbf {\bibinfo {volume} {99}},\ \bibinfo {pages} {014317} (\bibinfo {year} {2019})}\BibitemShut {NoStop}%
\bibitem [{\citenamefont {Manjunatha}\ \emph {et~al.}(2020)\citenamefont {Manjunatha}, \citenamefont {Sridhar}, \citenamefont {Damodara~Gupta}, \citenamefont {Ramalingam},\ and\ \citenamefont {Doddamani}}]{17}%
  \BibitemOpen
  \bibfield  {author} {\bibinfo {author} {\bibfnamefont {H.~C.}\ \bibnamefont {Manjunatha}}, \bibinfo {author} {\bibfnamefont {G.~R.}\ \bibnamefont {Sridhar}}, \bibinfo {author} {\bibfnamefont {P.~S.}\ \bibnamefont {Damodara~Gupta}}, \bibinfo {author} {\bibfnamefont {H.~B.}\ \bibnamefont {Ramalingam}},\ and\ \bibinfo {author} {\bibfnamefont {V.~H.}\ \bibnamefont {Doddamani}},\ }\bibfield  {title} {\bibinfo {title} {Pocket formula for alpha decay energies and half-lives of actinide nuclei},\ }\href@noop {} {\bibfield  {journal} {\bibinfo  {journal} {Zeitschrift f\"ur Naturforschung A}\ }\textbf {\bibinfo {volume} {75}},\ \bibinfo {pages} {501} (\bibinfo {year} {2020})}\BibitemShut {NoStop}%
\bibitem [{\citenamefont {Ding}\ and\ \citenamefont {Wan}(2026)}]{18}%
  \BibitemOpen
  \bibfield  {author} {\bibinfo {author} {\bibfnamefont {Y.-H.}\ \bibnamefont {Ding}}\ and\ \bibinfo {author} {\bibfnamefont {N.}~\bibnamefont {Wan}},\ }\bibfield  {title} {\bibinfo {title} {Effects of nuclear masses from different mass models on the $\alpha$-decay properties of superheavy nuclei},\ }\href {https://doi.org/10.1088/1674-1137/ae31e0} {\bibfield  {journal} {\bibinfo  {journal} {Chinese Physics C}\ }\textbf {\bibinfo {volume} {50}},\ \bibinfo {pages} {044109} (\bibinfo {year} {2026})}\BibitemShut {NoStop}%
\bibitem [{\citenamefont {Moller}\ \emph {et~al.}(2016)\citenamefont {Moller}, \citenamefont {Sierk}, \citenamefont {Ichikawa},\ and\ \citenamefont {Sagawa}}]{19}%
  \BibitemOpen
  \bibfield  {author} {\bibinfo {author} {\bibfnamefont {P.}~\bibnamefont {Moller}}, \bibinfo {author} {\bibfnamefont {A.~J.}\ \bibnamefont {Sierk}}, \bibinfo {author} {\bibfnamefont {T.}~\bibnamefont {Ichikawa}},\ and\ \bibinfo {author} {\bibfnamefont {H.}~\bibnamefont {Sagawa}},\ }\bibfield  {title} {\bibinfo {title} {Nuclear ground-state masses and deformations: {{FRDM}}(2012)},\ }\href@noop {} {\bibfield  {journal} {\bibinfo  {journal} {Atomic Data and Nuclear Data Tables}\ }\textbf {\bibinfo {volume} {109--110}},\ \bibinfo {pages} {1} (\bibinfo {year} {2016})}\BibitemShut {NoStop}%
\bibitem [{\citenamefont {Deng}\ \emph {et~al.}(2025{\natexlab{a}})\citenamefont {Deng}, \citenamefont {Li}, \citenamefont {Cheng}, \citenamefont {Mu},\ and\ \citenamefont {Zhang}}]{20}%
  \BibitemOpen
  \bibfield  {author} {\bibinfo {author} {\bibfnamefont {J.-G.}\ \bibnamefont {Deng}}, \bibinfo {author} {\bibfnamefont {J.-X.}\ \bibnamefont {Li}}, \bibinfo {author} {\bibfnamefont {J.-H.}\ \bibnamefont {Cheng}}, \bibinfo {author} {\bibfnamefont {L.}~\bibnamefont {Mu}},\ and\ \bibinfo {author} {\bibfnamefont {H.-F.}\ \bibnamefont {Zhang}},\ }\bibfield  {title} {\bibinfo {title} {\(\alpha\) decay properties of superheavy nuclei based on optimized \(\alpha\) decay energies},\ }\href {https://doi.org/10.1088/1674-1137/adb2fa} {\bibfield  {journal} {\bibinfo  {journal} {Chinese Physics C}\ }\textbf {\bibinfo {volume} {49}},\ \bibinfo {pages} {054104} (\bibinfo {year} {2025}{\natexlab{a}})}\BibitemShut {NoStop}%
\bibitem [{\citenamefont {Wang}\ \emph {et~al.}(2014)\citenamefont {Wang}, \citenamefont {Liu}, \citenamefont {Wu},\ and\ \citenamefont {Meng}}]{21}%
  \BibitemOpen
  \bibfield  {author} {\bibinfo {author} {\bibfnamefont {N.}~\bibnamefont {Wang}}, \bibinfo {author} {\bibfnamefont {M.}~\bibnamefont {Liu}}, \bibinfo {author} {\bibfnamefont {X.}~\bibnamefont {Wu}},\ and\ \bibinfo {author} {\bibfnamefont {J.}~\bibnamefont {Meng}},\ }\bibfield  {title} {\bibinfo {title} {Surface diffuseness correction in global mass formula},\ }\href@noop {} {\bibfield  {journal} {\bibinfo  {journal} {Physics Letters B}\ }\textbf {\bibinfo {volume} {734}},\ \bibinfo {pages} {215} (\bibinfo {year} {2014})}\BibitemShut {NoStop}%
\bibitem [{\citenamefont {Ring}(1996)}]{22}%
  \BibitemOpen
  \bibfield  {author} {\bibinfo {author} {\bibfnamefont {P.}~\bibnamefont {Ring}},\ }\bibfield  {title} {\bibinfo {title} {Relativistic mean field theory in finite nuclei},\ }\href@noop {} {\bibfield  {journal} {\bibinfo  {journal} {Progress in Particle and Nuclear Physics}\ }\textbf {\bibinfo {volume} {37}},\ \bibinfo {pages} {193} (\bibinfo {year} {1996})}\BibitemShut {NoStop}%
\bibitem [{\citenamefont {Grams}\ \emph {et~al.}(2026)\citenamefont {Grams}, \citenamefont {Ryssens}, \citenamefont {S{\'a}nchez-Fern{\'a}ndez}, \citenamefont {Shchechilin}, \citenamefont {Gonz{\'a}lez-Miret~Zaragoza}, \citenamefont {Demol}, \citenamefont {Chamel}, \citenamefont {Goriely},\ and\ \citenamefont {Bender}}]{23}%
  \BibitemOpen
  \bibfield  {author} {\bibinfo {author} {\bibfnamefont {G.}~\bibnamefont {Grams}}, \bibinfo {author} {\bibfnamefont {W.}~\bibnamefont {Ryssens}}, \bibinfo {author} {\bibfnamefont {A.}~\bibnamefont {S{\'a}nchez-Fern{\'a}ndez}}, \bibinfo {author} {\bibfnamefont {N.~N.}\ \bibnamefont {Shchechilin}}, \bibinfo {author} {\bibfnamefont {L.}~\bibnamefont {Gonz{\'a}lez-Miret~Zaragoza}}, \bibinfo {author} {\bibfnamefont {P.}~\bibnamefont {Demol}}, \bibinfo {author} {\bibfnamefont {N.}~\bibnamefont {Chamel}}, \bibinfo {author} {\bibfnamefont {S.}~\bibnamefont {Goriely}},\ and\ \bibinfo {author} {\bibfnamefont {M.}~\bibnamefont {Bender}},\ }\bibfield  {title} {\bibinfo {title} {Skyrme-hartree-fock-bogoliubov mass models on a 3d mesh: V. the n2lo extension of the skyrme edf},\ }\href {https://doi.org/10.1016/j.physletb.2026.140590} {\bibfield  {journal} {\bibinfo  {journal} {Physics Letters B}\ }\textbf {\bibinfo {volume} {879}},\ \bibinfo {pages} {140590} (\bibinfo {year} {2026})}\BibitemShut {NoStop}%
\bibitem [{\citenamefont {Goriely}\ \emph {et~al.}(2013)\citenamefont {Goriely}, \citenamefont {Chamel},\ and\ \citenamefont {Pearson}}]{24}%
  \BibitemOpen
  \bibfield  {author} {\bibinfo {author} {\bibfnamefont {S.}~\bibnamefont {Goriely}}, \bibinfo {author} {\bibfnamefont {N.}~\bibnamefont {Chamel}},\ and\ \bibinfo {author} {\bibfnamefont {J.~M.}\ \bibnamefont {Pearson}},\ }\bibfield  {title} {\bibinfo {title} {Further explorations of {{Skyrme-Hartree-Fock-Bogoliubov}} mass formulas. . {{The}} 2012 atomic mass evaluation and the symmetry coefficient},\ }\href@noop {} {\bibfield  {journal} {\bibinfo  {journal} {Physical Review C}\ }\textbf {\bibinfo {volume} {88}},\ \bibinfo {pages} {024308} (\bibinfo {year} {2013})}\BibitemShut {NoStop}%
\bibitem [{\citenamefont {Bhagwat}(2014)}]{25}%
  \BibitemOpen
  \bibfield  {author} {\bibinfo {author} {\bibfnamefont {A.}~\bibnamefont {Bhagwat}},\ }\bibfield  {title} {\bibinfo {title} {Simple nuclear mass formula},\ }\href@noop {} {\bibfield  {journal} {\bibinfo  {journal} {Physical Review C}\ }\textbf {\bibinfo {volume} {90}},\ \bibinfo {pages} {064306} (\bibinfo {year} {2014})}\BibitemShut {NoStop}%
\bibitem [{\citenamefont {Koura}\ \emph {et~al.}(2005)\citenamefont {Koura}, \citenamefont {Tachibana}, \citenamefont {Uno},\ and\ \citenamefont {Yamada}}]{26}%
  \BibitemOpen
  \bibfield  {author} {\bibinfo {author} {\bibfnamefont {H.}~\bibnamefont {Koura}}, \bibinfo {author} {\bibfnamefont {T.}~\bibnamefont {Tachibana}}, \bibinfo {author} {\bibfnamefont {M.}~\bibnamefont {Uno}},\ and\ \bibinfo {author} {\bibfnamefont {M.}~\bibnamefont {Yamada}},\ }\bibfield  {title} {\bibinfo {title} {Nuclidic {{Mass Formula}} on a {{Spherical Basis}} with an {{Improved Even-Odd Term}}},\ }\href@noop {} {\bibfield  {journal} {\bibinfo  {journal} {Progress of Theoretical Physics}\ }\textbf {\bibinfo {volume} {113}},\ \bibinfo {pages} {305} (\bibinfo {year} {2005})}\BibitemShut {NoStop}%
\bibitem [{\citenamefont {Duflo}\ and\ \citenamefont {Zuker}(1995)}]{27}%
  \BibitemOpen
  \bibfield  {author} {\bibinfo {author} {\bibfnamefont {J.}~\bibnamefont {Duflo}}\ and\ \bibinfo {author} {\bibfnamefont {A.~P.}\ \bibnamefont {Zuker}},\ }\bibfield  {title} {\bibinfo {title} {Microscopic mass formulae},\ }\href@noop {} {\bibfield  {journal} {\bibinfo  {journal} {Physical Review C}\ }\textbf {\bibinfo {volume} {52}},\ \bibinfo {pages} {R23} (\bibinfo {year} {1995})}\BibitemShut {NoStop}%
\bibitem [{\citenamefont {Lu}\ \emph {et~al.}(2025)\citenamefont {Lu}, \citenamefont {Shang}, \citenamefont {Du}, \citenamefont {Li}, \citenamefont {Liang},\ and\ \citenamefont {Niu}}]{28}%
  \BibitemOpen
  \bibfield  {author} {\bibinfo {author} {\bibfnamefont {Y.}~\bibnamefont {Lu}}, \bibinfo {author} {\bibfnamefont {T.}~\bibnamefont {Shang}}, \bibinfo {author} {\bibfnamefont {P.}~\bibnamefont {Du}}, \bibinfo {author} {\bibfnamefont {J.}~\bibnamefont {Li}}, \bibinfo {author} {\bibfnamefont {H.}~\bibnamefont {Liang}},\ and\ \bibinfo {author} {\bibfnamefont {Z.}~\bibnamefont {Niu}},\ }\bibfield  {title} {\bibinfo {title} {Nuclear mass predictions based on convolutional neural network},\ }\href@noop {} {\bibfield  {journal} {\bibinfo  {journal} {Physical Review C}\ }\textbf {\bibinfo {volume} {111}},\ \bibinfo {pages} {014325} (\bibinfo {year} {2025})}\BibitemShut {NoStop}%
\bibitem [{\citenamefont {Wang}\ \emph {et~al.}(2016)\citenamefont {Wang}, \citenamefont {Liu}, \citenamefont {Wu},\ and\ \citenamefont {Meng}}]{29}%
  \BibitemOpen
  \bibfield  {author} {\bibinfo {author} {\bibfnamefont {N.}~\bibnamefont {Wang}}, \bibinfo {author} {\bibfnamefont {M.}~\bibnamefont {Liu}}, \bibinfo {author} {\bibfnamefont {X.}~\bibnamefont {Wu}},\ and\ \bibinfo {author} {\bibfnamefont {J.}~\bibnamefont {Meng}},\ }\bibfield  {title} {\bibinfo {title} {Correlations between neutrons and protons near {{Fermi}} surface and {$Q_{\alpha}$} of super-heavy nuclei},\ }\href@noop {} {\bibfield  {journal} {\bibinfo  {journal} {Physical Review C}\ }\textbf {\bibinfo {volume} {93}},\ \bibinfo {pages} {014302} (\bibinfo {year} {2016})}\BibitemShut {NoStop}%
\bibitem [{\citenamefont {Liu}\ \emph {et~al.}(2011)\citenamefont {Liu}, \citenamefont {Wang}, \citenamefont {Deng},\ and\ \citenamefont {Wu}}]{30}%
  \BibitemOpen
  \bibfield  {author} {\bibinfo {author} {\bibfnamefont {M.}~\bibnamefont {Liu}}, \bibinfo {author} {\bibfnamefont {N.}~\bibnamefont {Wang}}, \bibinfo {author} {\bibfnamefont {Y.}~\bibnamefont {Deng}},\ and\ \bibinfo {author} {\bibfnamefont {X.}~\bibnamefont {Wu}},\ }\bibfield  {title} {\bibinfo {title} {Further improvements on a global nuclear mass model},\ }\href@noop {} {\bibfield  {journal} {\bibinfo  {journal} {Physical Review C}\ }\textbf {\bibinfo {volume} {84}},\ \bibinfo {pages} {014333} (\bibinfo {year} {2011})}\BibitemShut {NoStop}%
\bibitem [{\citenamefont {Ma}\ \emph {et~al.}(2019)\citenamefont {Ma}, \citenamefont {Zhang}, \citenamefont {Bao},\ and\ \citenamefont {Zhang}}]{31}%
  \BibitemOpen
  \bibfield  {author} {\bibinfo {author} {\bibfnamefont {N.-N.}\ \bibnamefont {Ma}}, \bibinfo {author} {\bibfnamefont {H.-F.}\ \bibnamefont {Zhang}}, \bibinfo {author} {\bibfnamefont {X.-J.}\ \bibnamefont {Bao}},\ and\ \bibinfo {author} {\bibfnamefont {H.-F.}\ \bibnamefont {Zhang}},\ }\bibfield  {title} {\bibinfo {title} {Basic characteristics of nuclear landscape by improved {{Weizs\"acker-Skyrme}}-type nuclear mass model},\ }\href@noop {} {\bibfield  {journal} {\bibinfo  {journal} {Chinese Physics C}\ }\textbf {\bibinfo {volume} {43}},\ \bibinfo {pages} {044105} (\bibinfo {year} {2019})}\BibitemShut {NoStop}%
\bibitem [{\citenamefont {Li}\ \emph {et~al.}(2021)\citenamefont {Li}, \citenamefont {Wei}, \citenamefont {Liu},\ and\ \citenamefont {Wang}}]{32}%
  \BibitemOpen
  \bibfield  {author} {\bibinfo {author} {\bibfnamefont {T.}~\bibnamefont {Li}}, \bibinfo {author} {\bibfnamefont {H.}~\bibnamefont {Wei}}, \bibinfo {author} {\bibfnamefont {M.}~\bibnamefont {Liu}},\ and\ \bibinfo {author} {\bibfnamefont {N.}~\bibnamefont {Wang}},\ }\bibfield  {title} {\bibinfo {title} {Ability of the radial basis function approach to extrapolate nuclear mass},\ }\href@noop {} {\bibfield  {journal} {\bibinfo  {journal} {Communications in Theoretical Physics}\ }\textbf {\bibinfo {volume} {73}},\ \bibinfo {pages} {095301} (\bibinfo {year} {2021})}\BibitemShut {NoStop}%
\bibitem [{\citenamefont {Niu}\ and\ \citenamefont {Liang}(2018)}]{33}%
  \BibitemOpen
  \bibfield  {author} {\bibinfo {author} {\bibfnamefont {Z.~M.}\ \bibnamefont {Niu}}\ and\ \bibinfo {author} {\bibfnamefont {H.~Z.}\ \bibnamefont {Liang}},\ }\bibfield  {title} {\bibinfo {title} {Nuclear mass predictions based on {{Bayesian}} neural network approach with pairing and shell effects},\ }\href@noop {} {\bibfield  {journal} {\bibinfo  {journal} {Physics Letters B}\ }\textbf {\bibinfo {volume} {778}},\ \bibinfo {pages} {48} (\bibinfo {year} {2018})}\BibitemShut {NoStop}%
\bibitem [{\citenamefont {Gao}\ \emph {et~al.}(2021)\citenamefont {Gao}, \citenamefont {Wang}, \citenamefont {L{\"u}}, \citenamefont {Li}, \citenamefont {Shen},\ and\ \citenamefont {Liu}}]{34}%
  \BibitemOpen
  \bibfield  {author} {\bibinfo {author} {\bibfnamefont {Z.}~\bibnamefont {Gao}}, \bibinfo {author} {\bibfnamefont {Y.}~\bibnamefont {Wang}}, \bibinfo {author} {\bibfnamefont {H.}~\bibnamefont {L{\"u}}}, \bibinfo {author} {\bibfnamefont {Q.}~\bibnamefont {Li}}, \bibinfo {author} {\bibfnamefont {C.}~\bibnamefont {Shen}},\ and\ \bibinfo {author} {\bibfnamefont {L.}~\bibnamefont {Liu}},\ }\bibfield  {title} {\bibinfo {title} {Machine learning the nuclear mass},\ }\href@noop {} {\bibfield  {journal} {\bibinfo  {journal} {Nuclear Science and Techniques}\ }\textbf {\bibinfo {volume} {32}},\ \bibinfo {pages} {109} (\bibinfo {year} {2021})}\BibitemShut {NoStop}%
\bibitem [{\citenamefont {Athanassopoulos}\ \emph {et~al.}(2004)\citenamefont {Athanassopoulos}, \citenamefont {Mavrommatis}, \citenamefont {Gernoth},\ and\ \citenamefont {Clark}}]{35}%
  \BibitemOpen
  \bibfield  {author} {\bibinfo {author} {\bibfnamefont {S.}~\bibnamefont {Athanassopoulos}}, \bibinfo {author} {\bibfnamefont {E.}~\bibnamefont {Mavrommatis}}, \bibinfo {author} {\bibfnamefont {K.~A.}\ \bibnamefont {Gernoth}},\ and\ \bibinfo {author} {\bibfnamefont {J.~W.}\ \bibnamefont {Clark}},\ }\bibfield  {title} {\bibinfo {title} {Nuclear mass systematics using neural networks},\ }\href@noop {} {\bibfield  {journal} {\bibinfo  {journal} {Nuclear Physics A}\ }\textbf {\bibinfo {volume} {743}},\ \bibinfo {pages} {222} (\bibinfo {year} {2004})}\BibitemShut {NoStop}%
\bibitem [{\citenamefont {Wu}\ \emph {et~al.}(2020)\citenamefont {Wu}, \citenamefont {Bai}, \citenamefont {Sagawa},\ and\ \citenamefont {Zhang}}]{36}%
  \BibitemOpen
  \bibfield  {author} {\bibinfo {author} {\bibfnamefont {D.}~\bibnamefont {Wu}}, \bibinfo {author} {\bibfnamefont {C.~L.}\ \bibnamefont {Bai}}, \bibinfo {author} {\bibfnamefont {H.}~\bibnamefont {Sagawa}},\ and\ \bibinfo {author} {\bibfnamefont {H.~Q.}\ \bibnamefont {Zhang}},\ }\bibfield  {title} {\bibinfo {title} {Study of {{Charge Radii}} with {{Neural Networks}}},\ }\href@noop {} {\bibfield  {journal} {\bibinfo  {journal} {Physical Review C}\ }\textbf {\bibinfo {volume} {102}},\ \bibinfo {pages} {054323} (\bibinfo {year} {2020})}\BibitemShut {NoStop}%
\bibitem [{\citenamefont {Richardson}\ \emph {et~al.}(2026)\citenamefont {Richardson}, \citenamefont {Trifinopoulos},\ and\ \citenamefont {Williams}}]{37}%
  \BibitemOpen
  \bibfield  {author} {\bibinfo {author} {\bibfnamefont {K.~A.}\ \bibnamefont {Richardson}}, \bibinfo {author} {\bibfnamefont {S.}~\bibnamefont {Trifinopoulos}},\ and\ \bibinfo {author} {\bibfnamefont {M.}~\bibnamefont {Williams}},\ }\bibfield  {title} {\bibinfo {title} {The {DNA} of nuclear models: How {AI} predicts nuclear masses},\ }\href {https://doi.org/10.48550/arXiv.2508.08370} {\bibfield  {journal} {\bibinfo  {journal} {Communications AI \& Computing}\ }\textbf {\bibinfo {volume} {1}},\ \bibinfo {pages} {20} (\bibinfo {year} {2026})}\BibitemShut {NoStop}%
\bibitem [{\citenamefont {Zhao}\ and\ \citenamefont {Bao}(2026)}]{38}%
  \BibitemOpen
  \bibfield  {author} {\bibinfo {author} {\bibfnamefont {T.~L.}\ \bibnamefont {Zhao}}\ and\ \bibinfo {author} {\bibfnamefont {X.~J.}\ \bibnamefont {Bao}},\ }\bibfield  {title} {\bibinfo {title} {Explore the high precision and interpretability of deep learning in the calculation of {$\alpha$} decay half life},\ }\href@noop {} {\bibfield  {journal} {\bibinfo  {journal} {Physics Letters B}\ }\textbf {\bibinfo {volume} {879}},\ \bibinfo {pages} {140609} (\bibinfo {year} {2026})}\BibitemShut {NoStop}%
\bibitem [{\citenamefont {Y{\"u}ksel}\ \emph {et~al.}(2024)\citenamefont {Y{\"u}ksel}, \citenamefont {Soydaner},\ and\ \citenamefont {Bahtiyar}}]{39}%
  \BibitemOpen
  \bibfield  {author} {\bibinfo {author} {\bibfnamefont {E.}~\bibnamefont {Y{\"u}ksel}}, \bibinfo {author} {\bibfnamefont {D.}~\bibnamefont {Soydaner}},\ and\ \bibinfo {author} {\bibfnamefont {H.}~\bibnamefont {Bahtiyar}},\ }\bibfield  {title} {\bibinfo {title} {Nuclear mass predictions using machine learning models},\ }\href@noop {} {\bibfield  {journal} {\bibinfo  {journal} {Physical Review C}\ }\textbf {\bibinfo {volume} {109}},\ \bibinfo {pages} {064322} (\bibinfo {year} {2024})}\BibitemShut {NoStop}%
\bibitem [{\citenamefont {Li}\ \emph {et~al.}(2022)\citenamefont {Li}, \citenamefont {Tong}, \citenamefont {Du},\ and\ \citenamefont {Pang}}]{40}%
  \BibitemOpen
  \bibfield  {author} {\bibinfo {author} {\bibfnamefont {C.-Q.}\ \bibnamefont {Li}}, \bibinfo {author} {\bibfnamefont {C.-N.}\ \bibnamefont {Tong}}, \bibinfo {author} {\bibfnamefont {H.-J.}\ \bibnamefont {Du}},\ and\ \bibinfo {author} {\bibfnamefont {L.-G.}\ \bibnamefont {Pang}},\ }\bibfield  {title} {\bibinfo {title} {Deep learning on nuclear mass and {$\alpha$} decay half-lives},\ }\href@noop {} {\bibfield  {journal} {\bibinfo  {journal} {Physical Review C}\ }\textbf {\bibinfo {volume} {105}},\ \bibinfo {pages} {064306} (\bibinfo {year} {2022})}\BibitemShut {NoStop}%
\bibitem [{\citenamefont {Cai}\ and\ \citenamefont {Yuan}(2023)}]{41}%
  \BibitemOpen
  \bibfield  {author} {\bibinfo {author} {\bibfnamefont {B.}~\bibnamefont {Cai}}\ and\ \bibinfo {author} {\bibfnamefont {C.}~\bibnamefont {Yuan}},\ }\bibfield  {title} {\bibinfo {title} {Decay of superheavy nuclei based on the random forest algorithm},\ }\href@noop {} {\bibfield  {journal} {\bibinfo  {journal} {Nuclear Science and Techniques}\ }\textbf {\bibinfo {volume} {34}},\ \bibinfo {pages} {204} (\bibinfo {year} {2023})}\BibitemShut {NoStop}%
\bibitem [{\citenamefont {Yuan}\ \emph {et~al.}(2026)\citenamefont {Yuan}, \citenamefont {Xiao}, \citenamefont {Qi}, \citenamefont {Yang}, \citenamefont {Yu}, \citenamefont {Yang}, \citenamefont {Li},\ and\ \citenamefont {Cai}}]{42}%
  \BibitemOpen
  \bibfield  {author} {\bibinfo {author} {\bibfnamefont {Q.}~\bibnamefont {Yuan}}, \bibinfo {author} {\bibfnamefont {X.}~\bibnamefont {Xiao}}, \bibinfo {author} {\bibfnamefont {P.}~\bibnamefont {Qi}}, \bibinfo {author} {\bibfnamefont {A.}~\bibnamefont {Yang}}, \bibinfo {author} {\bibfnamefont {G.}~\bibnamefont {Yu}}, \bibinfo {author} {\bibfnamefont {H.}~\bibnamefont {Yang}}, \bibinfo {author} {\bibfnamefont {Z.}~\bibnamefont {Li}},\ and\ \bibinfo {author} {\bibfnamefont {Y.}~\bibnamefont {Cai}},\ }\bibfield  {title} {\bibinfo {title} {Physics-guided residual correction of {\ensuremath{\alpha}}-decay half-lives based on the effective liquid drop model},\ }\href {https://doi.org/10.1103/7gm3-2y7f} {\bibfield  {journal} {\bibinfo  {journal} {Physical Review C}\ }\textbf {\bibinfo {volume} {114}},\ \bibinfo {pages} {044625} (\bibinfo {year} {2026})}\BibitemShut {NoStop}%
\bibitem [{\citenamefont {Lu}\ \emph {et~al.}(2026)\citenamefont {Lu}, \citenamefont {Shang}, \citenamefont {Du}, \citenamefont {Li},\ and\ \citenamefont {Liang}}]{43}%
  \BibitemOpen
  \bibfield  {author} {\bibinfo {author} {\bibfnamefont {Y.}~\bibnamefont {Lu}}, \bibinfo {author} {\bibfnamefont {T.}~\bibnamefont {Shang}}, \bibinfo {author} {\bibfnamefont {P.}~\bibnamefont {Du}}, \bibinfo {author} {\bibfnamefont {J.}~\bibnamefont {Li}},\ and\ \bibinfo {author} {\bibfnamefont {H.}~\bibnamefont {Liang}},\ }\bibfield  {title} {\bibinfo {title} {Correcting nuclear mass models with interpretable machine learning to bridge theory and data},\ }\href {https://doi.org/10.1103/yjnd-vgjr} {\bibfield  {journal} {\bibinfo  {journal} {Physical Review C}\ }\textbf {\bibinfo {volume} {114}},\ \bibinfo {pages} {034310} (\bibinfo {year} {2026})}\BibitemShut {NoStop}%
\bibitem [{\citenamefont {McCulloch}\ and\ \citenamefont {Pitts}(1943)}]{44}%
  \BibitemOpen
  \bibfield  {author} {\bibinfo {author} {\bibfnamefont {W.~S.}\ \bibnamefont {McCulloch}}\ and\ \bibinfo {author} {\bibfnamefont {W.}~\bibnamefont {Pitts}},\ }\bibfield  {title} {\bibinfo {title} {A logical calculus of the ideas immanent in nervous activity},\ }\href@noop {} {\bibfield  {journal} {\bibinfo  {journal} {The Bulletin of Mathematical Biophysics}\ }\textbf {\bibinfo {volume} {5}},\ \bibinfo {pages} {115} (\bibinfo {year} {1943})}\BibitemShut {NoStop}%
\bibitem [{\citenamefont {Cortes}\ and\ \citenamefont {Vapnik}(1995)}]{45}%
  \BibitemOpen
  \bibfield  {author} {\bibinfo {author} {\bibfnamefont {C.}~\bibnamefont {Cortes}}\ and\ \bibinfo {author} {\bibfnamefont {V.}~\bibnamefont {Vapnik}},\ }\bibfield  {title} {\bibinfo {title} {Support-vector networks},\ }\href@noop {} {\bibfield  {journal} {\bibinfo  {journal} {Machine Learning}\ }\textbf {\bibinfo {volume} {20}},\ \bibinfo {pages} {273} (\bibinfo {year} {1995})}\BibitemShut {NoStop}%
\bibitem [{\citenamefont {Breiman}(2001)}]{46}%
  \BibitemOpen
  \bibfield  {author} {\bibinfo {author} {\bibfnamefont {L.}~\bibnamefont {Breiman}},\ }\bibfield  {title} {\bibinfo {title} {Random forests},\ }\href {https://doi.org/10.1023/A:1010933404324} {\bibfield  {journal} {\bibinfo  {journal} {Machine Learning}\ }\textbf {\bibinfo {volume} {45}},\ \bibinfo {pages} {5} (\bibinfo {year} {2001})}\BibitemShut {NoStop}%
\bibitem [{\citenamefont {Shahaf}\ \emph {et~al.}(2015)\citenamefont {Shahaf}, \citenamefont {Guestrin}, \citenamefont {Horvitz},\ and\ \citenamefont {Leskovec}}]{47}%
  \BibitemOpen
  \bibfield  {author} {\bibinfo {author} {\bibfnamefont {D.}~\bibnamefont {Shahaf}}, \bibinfo {author} {\bibfnamefont {C.}~\bibnamefont {Guestrin}}, \bibinfo {author} {\bibfnamefont {E.}~\bibnamefont {Horvitz}},\ and\ \bibinfo {author} {\bibfnamefont {J.}~\bibnamefont {Leskovec}},\ }\bibfield  {title} {\bibinfo {title} {Information cartography},\ }\href {https://doi.org/10.1145/2735624} {\bibfield  {journal} {\bibinfo  {journal} {Communications of the ACM}\ }\textbf {\bibinfo {volume} {58}},\ \bibinfo {pages} {68} (\bibinfo {year} {2015})}\BibitemShut {NoStop}%
\bibitem [{\citenamefont {Hollmann}\ \emph {et~al.}(2025)\citenamefont {Hollmann}, \citenamefont {M{\"u}ller}, \citenamefont {Purucker}, \citenamefont {Krishnakumar}, \citenamefont {K{\"o}rfer}, \citenamefont {Hoo}, \citenamefont {Schirrmeister},\ and\ \citenamefont {Hutter}}]{48}%
  \BibitemOpen
  \bibfield  {author} {\bibinfo {author} {\bibfnamefont {N.}~\bibnamefont {Hollmann}}, \bibinfo {author} {\bibfnamefont {S.}~\bibnamefont {M{\"u}ller}}, \bibinfo {author} {\bibfnamefont {L.}~\bibnamefont {Purucker}}, \bibinfo {author} {\bibfnamefont {A.}~\bibnamefont {Krishnakumar}}, \bibinfo {author} {\bibfnamefont {M.}~\bibnamefont {K{\"o}rfer}}, \bibinfo {author} {\bibfnamefont {S.~B.}\ \bibnamefont {Hoo}}, \bibinfo {author} {\bibfnamefont {R.~T.}\ \bibnamefont {Schirrmeister}},\ and\ \bibinfo {author} {\bibfnamefont {F.}~\bibnamefont {Hutter}},\ }\bibfield  {title} {\bibinfo {title} {Accurate predictions on small data with a tabular foundation model},\ }\href@noop {} {\bibfield  {journal} {\bibinfo  {journal} {Nature}\ }\textbf {\bibinfo {volume} {637}},\ \bibinfo {pages} {319} (\bibinfo {year} {2025})}\BibitemShut {NoStop}%
\bibitem [{\citenamefont {Qi}\ \emph {et~al.}(2026)\citenamefont {Qi}, \citenamefont {Xiao}, \citenamefont {Yu}, \citenamefont {Yang},\ and\ \citenamefont {Hu}}]{49}%
  \BibitemOpen
  \bibfield  {author} {\bibinfo {author} {\bibfnamefont {P.}~\bibnamefont {Qi}}, \bibinfo {author} {\bibfnamefont {X.}~\bibnamefont {Xiao}}, \bibinfo {author} {\bibfnamefont {G.}~\bibnamefont {Yu}}, \bibinfo {author} {\bibfnamefont {H.}~\bibnamefont {Yang}},\ and\ \bibinfo {author} {\bibfnamefont {Q.}~\bibnamefont {Hu}},\ }\bibfield  {title} {\bibinfo {title} {Systematic study on the {$\alpha$}-particle preformation factor in the theory of {$\alpha$}-decay based on the tabular prior-data fitted network ({TabPFN})},\ }\href {https://doi.org/10.1103/5v47-6dcs} {\bibfield  {journal} {\bibinfo  {journal} {Physical Review C}\ }\textbf {\bibinfo {volume} {113}},\ \bibinfo {pages} {054606} (\bibinfo {year} {2026})}\BibitemShut {NoStop}%
\bibitem [{\citenamefont {Wang}\ and\ \citenamefont {Liu}(2011)}]{50}%
  \BibitemOpen
  \bibfield  {author} {\bibinfo {author} {\bibfnamefont {N.}~\bibnamefont {Wang}}\ and\ \bibinfo {author} {\bibfnamefont {M.}~\bibnamefont {Liu}},\ }\bibfield  {title} {\bibinfo {title} {Nuclear mass predictions with radial basis function approach},\ }\href@noop {} {\bibfield  {journal} {\bibinfo  {journal} {Physical Review C}\ }\textbf {\bibinfo {volume} {84}},\ \bibinfo {pages} {051303} (\bibinfo {year} {2011})}\BibitemShut {NoStop}%
\bibitem [{\citenamefont {Cai}\ \emph {et~al.}(2020)\citenamefont {Cai}, \citenamefont {Chen}, \citenamefont {Xu}, \citenamefont {Yuan}, \citenamefont {Qi},\ and\ \citenamefont {Yao}}]{51}%
  \BibitemOpen
  \bibfield  {author} {\bibinfo {author} {\bibfnamefont {B.}~\bibnamefont {Cai}}, \bibinfo {author} {\bibfnamefont {G.}~\bibnamefont {Chen}}, \bibinfo {author} {\bibfnamefont {J.}~\bibnamefont {Xu}}, \bibinfo {author} {\bibfnamefont {C.}~\bibnamefont {Yuan}}, \bibinfo {author} {\bibfnamefont {C.}~\bibnamefont {Qi}},\ and\ \bibinfo {author} {\bibfnamefont {Y.}~\bibnamefont {Yao}},\ }\bibfield  {title} {\bibinfo {title} {{$\alpha$} decay half-life estimation and uncertainty analysis},\ }\href@noop {} {\bibfield  {journal} {\bibinfo  {journal} {Physical Review C}\ }\textbf {\bibinfo {volume} {101}},\ \bibinfo {pages} {054304} (\bibinfo {year} {2020})}\BibitemShut {NoStop}%
\bibitem [{\citenamefont {Cai}\ \emph {et~al.}(2026)\citenamefont {Cai}, \citenamefont {Yuan},\ and\ \citenamefont {Qi}}]{52}%
  \BibitemOpen
  \bibfield  {author} {\bibinfo {author} {\bibfnamefont {B.}~\bibnamefont {Cai}}, \bibinfo {author} {\bibfnamefont {C.}~\bibnamefont {Yuan}},\ and\ \bibinfo {author} {\bibfnamefont {C.}~\bibnamefont {Qi}},\ }\bibfield  {title} {\bibinfo {title} {A universal four-fermion formation framework and odd-even staggering in {$\alpha$} decay},\ }\href {https://doi.org/10.1007/s11433-025-2911-9} {\bibfield  {journal} {\bibinfo  {journal} {Science China Physics, Mechanics \& Astronomy}\ }\textbf {\bibinfo {volume} {69}},\ \bibinfo {pages} {252011} (\bibinfo {year} {2026})}\BibitemShut {NoStop}%
\bibitem [{\citenamefont {Kirson}(2008)}]{53}%
  \BibitemOpen
  \bibfield  {author} {\bibinfo {author} {\bibfnamefont {M.~W.}\ \bibnamefont {Kirson}},\ }\bibfield  {title} {\bibinfo {title} {Mutual influence of terms in a semi-empirical mass formula},\ }\href@noop {} {\bibfield  {journal} {\bibinfo  {journal} {Nuclear Physics A}\ }\textbf {\bibinfo {volume} {798}},\ \bibinfo {pages} {29} (\bibinfo {year} {2008})}\BibitemShut {NoStop}%
\bibitem [{\citenamefont {Mo}\ \emph {et~al.}(2014)\citenamefont {Mo}, \citenamefont {Liu},\ and\ \citenamefont {Wang}}]{54}%
  \BibitemOpen
  \bibfield  {author} {\bibinfo {author} {\bibfnamefont {Q.}~\bibnamefont {Mo}}, \bibinfo {author} {\bibfnamefont {M.}~\bibnamefont {Liu}},\ and\ \bibinfo {author} {\bibfnamefont {N.}~\bibnamefont {Wang}},\ }\bibfield  {title} {\bibinfo {title} {Systematic study of shell gaps in nuclei},\ }\href@noop {} {\bibfield  {journal} {\bibinfo  {journal} {Physical Review C}\ }\textbf {\bibinfo {volume} {90}},\ \bibinfo {pages} {024320} (\bibinfo {year} {2014})}\BibitemShut {NoStop}%
\bibitem [{\citenamefont {Huang}\ \emph {et~al.}(2021)\citenamefont {Huang}, \citenamefont {Wang}, \citenamefont {Kondev}, \citenamefont {Audi},\ and\ \citenamefont {Naimi}}]{55}%
  \BibitemOpen
  \bibfield  {author} {\bibinfo {author} {\bibfnamefont {W.}~\bibnamefont {Huang}}, \bibinfo {author} {\bibfnamefont {M.}~\bibnamefont {Wang}}, \bibinfo {author} {\bibfnamefont {F.}~\bibnamefont {Kondev}}, \bibinfo {author} {\bibfnamefont {G.}~\bibnamefont {Audi}},\ and\ \bibinfo {author} {\bibfnamefont {S.}~\bibnamefont {Naimi}},\ }\bibfield  {title} {\bibinfo {title} {The {{AME}} 2020 atomic mass evaluation ({{I}}). {{Evaluation}} of input data, and adjustment procedures},\ }\href@noop {} {\bibfield  {journal} {\bibinfo  {journal} {Chinese Physics C}\ }\textbf {\bibinfo {volume} {45}},\ \bibinfo {pages} {030002} (\bibinfo {year} {2021})}\BibitemShut {NoStop}%
\bibitem [{\citenamefont {You}\ \emph {et~al.}(2022)\citenamefont {You}, \citenamefont {Qu}, \citenamefont {Wu}, \citenamefont {Su},\ and\ \citenamefont {He}}]{56}%
  \BibitemOpen
  \bibfield  {author} {\bibinfo {author} {\bibfnamefont {H.-Q.}\ \bibnamefont {You}}, \bibinfo {author} {\bibfnamefont {Z.-Z.}\ \bibnamefont {Qu}}, \bibinfo {author} {\bibfnamefont {R.-H.}\ \bibnamefont {Wu}}, \bibinfo {author} {\bibfnamefont {H.-Z.}\ \bibnamefont {Su}},\ and\ \bibinfo {author} {\bibfnamefont {X.-T.}\ \bibnamefont {He}},\ }\bibfield  {title} {\bibinfo {title} {Study of {$\alpha$}-decay energy by an artificial neural network considering pairing and shell effects},\ }\href {https://doi.org/10.3390/sym14051006} {\bibfield  {journal} {\bibinfo  {journal} {Symmetry}\ }\textbf {\bibinfo {volume} {14}},\ \bibinfo {pages} {1006} (\bibinfo {year} {2022})}\BibitemShut {NoStop}%
\bibitem [{\citenamefont {Ye}\ \emph {et~al.}(2024)\citenamefont {Ye}, \citenamefont {Zhou}, \citenamefont {Li},\ and\ \citenamefont {Zhan}}]{57}%
  \BibitemOpen
  \bibfield  {author} {\bibinfo {author} {\bibfnamefont {H.-J.}\ \bibnamefont {Ye}}, \bibinfo {author} {\bibfnamefont {D.-W.}\ \bibnamefont {Zhou}}, \bibinfo {author} {\bibfnamefont {L.}~\bibnamefont {Li}},\ and\ \bibinfo {author} {\bibfnamefont {D.-C.}\ \bibnamefont {Zhan}},\ }\bibfield  {title} {\bibinfo {title} {Contextualizing meta-learning via learning to decompose},\ }\href {https://doi.org/10.1109/TPAMI.2023.3317425} {\bibfield  {journal} {\bibinfo  {journal} {IEEE Transactions on Pattern Analysis and Machine Intelligence}\ }\textbf {\bibinfo {volume} {46}},\ \bibinfo {pages} {117} (\bibinfo {year} {2024})}\BibitemShut {NoStop}%
\bibitem [{\citenamefont {Deng}\ \emph {et~al.}(2025{\natexlab{b}})\citenamefont {Deng}, \citenamefont {Li}, \citenamefont {Cheng}, \citenamefont {Mu},\ and\ \citenamefont {Zhang}}]{59}%
  \BibitemOpen
  \bibfield  {author} {\bibinfo {author} {\bibfnamefont {J.-G.}\ \bibnamefont {Deng}}, \bibinfo {author} {\bibfnamefont {J.-X.}\ \bibnamefont {Li}}, \bibinfo {author} {\bibfnamefont {J.-H.}\ \bibnamefont {Cheng}}, \bibinfo {author} {\bibfnamefont {L.}~\bibnamefont {Mu}},\ and\ \bibinfo {author} {\bibfnamefont {H.-F.}\ \bibnamefont {Zhang}},\ }\bibfield  {title} {\bibinfo {title} {\(\alpha\) decay properties of superheavy nuclei based on optimized \(\alpha\) decay energies},\ }\href {https://doi.org/10.1088/1674-1137/adb2fa} {\bibfield  {journal} {\bibinfo  {journal} {Chinese Physics C}\ }\textbf {\bibinfo {volume} {49}},\ \bibinfo {pages} {054104} (\bibinfo {year} {2025}{\natexlab{b}})}\BibitemShut {NoStop}%
\bibitem [{\citenamefont {{\'C}wiok}\ \emph {et~al.}(1996)\citenamefont {{\'C}wiok}, \citenamefont {Dobaczewski}, \citenamefont {Heenen}, \citenamefont {Magierski},\ and\ \citenamefont {Nazarewicz}}]{60}%
  \BibitemOpen
  \bibfield  {author} {\bibinfo {author} {\bibfnamefont {S.}~\bibnamefont {{\'C}wiok}}, \bibinfo {author} {\bibfnamefont {J.}~\bibnamefont {Dobaczewski}}, \bibinfo {author} {\bibfnamefont {P.-H.}\ \bibnamefont {Heenen}}, \bibinfo {author} {\bibfnamefont {P.}~\bibnamefont {Magierski}},\ and\ \bibinfo {author} {\bibfnamefont {W.}~\bibnamefont {Nazarewicz}},\ }\bibfield  {title} {\bibinfo {title} {Shell {{Structure}} of the {{Superheavy Elements}}},\ }\href@noop {} {\bibfield  {journal} {\bibinfo  {journal} {Nuclear Physics A}\ }\textbf {\bibinfo {volume} {611}},\ \bibinfo {pages} {211} (\bibinfo {year} {1996})}\BibitemShut {NoStop}%
\bibitem [{\citenamefont {Jain}\ \emph {et~al.}(2024)\citenamefont {Jain}, \citenamefont {Bhuyan},\ and\ \citenamefont {Kumar}}]{58}%
  \BibitemOpen
  \bibfield  {author} {\bibinfo {author} {\bibfnamefont {N.}~\bibnamefont {Jain}}, \bibinfo {author} {\bibfnamefont {M.}~\bibnamefont {Bhuyan}},\ and\ \bibinfo {author} {\bibfnamefont {R.}~\bibnamefont {Kumar}},\ }\bibfield  {title} {\bibinfo {title} {Relativistic mean-field study of alpha decay in superheavy isotopes with 100 {$\leq$} z {$\leq$} 120},\ }\href {https://doi.org/10.1134/S1547477124700791} {\bibfield  {journal} {\bibinfo  {journal} {Physics of Particles and Nuclei Letters}\ }\textbf {\bibinfo {volume} {21}},\ \bibinfo {pages} {852} (\bibinfo {year} {2024})}\BibitemShut {NoStop}%
\end{thebibliography}%
\nocite{42,43,44,45,46,47,48,49,50,51,52,53,54,55,56,57,58,59,60}
\end{document}